\documentclass[aps,prb,twocolumn,showpacs]{revtex4-1}
\usepackage{mathptmx}

\usepackage[T1]{fontenc}
\usepackage[latin9]{inputenc}
\usepackage{float}
\usepackage{amsmath}
\usepackage{amssymb}
\usepackage{graphicx}
\usepackage{esint}
\usepackage[normalem]{ulem}

\makeatletter
\@ifundefined{textcolor}{}
{%
 \definecolor{BLACK}{gray}{0}
 \definecolor{WHITE}{gray}{1}
 \definecolor{RED}{rgb}{1,0,0}
 \definecolor{GREEN}{rgb}{0,1,0}
 \definecolor{BLUE}{rgb}{0,0,1}
 \definecolor{CYAN}{cmyk}{1,0,0,0}
 \definecolor{MAGENTA}{cmyk}{0,1,0,0}
 \definecolor{YELLOW}{cmyk}{0,0,1,0}
}

\pdfoutput=1 

\usepackage{multirow}\usepackage{dcolumn}\usepackage{bm}\usepackage{ulem}\usepackage{amsfonts}\usepackage{color}\usepackage{colordvi}\usepackage{dcolumn}\usepackage{subfigure}

\newcommand{\diag}{\text{diag}}
\newcommand{\tr}{\text{Tr}}

\usepackage[colorlinks,bookmarks=false,citecolor=blue,linkcolor=blue,urlcolor=blue,hyperfootnotes=true]{hyperref}

\makeatother

\begin{document}

\title{SU(2) symmetry in a realistic spin-fermion model for cuprate superconductors}

\date{May 22, 2015}

\author{T.\ Kloss$^{1,2}$, X.\ Montiel$^{1,2}$, C.\ P\'epin$^{1}$}

\affiliation{
$^{1}$IPhT, L'Orme des Merisiers, CEA-Saclay, 91191 Gif-sur-Yvette,
France }
\affiliation{
$^{2}$\mbox{IIP, Universidade Federal do Rio Grande do Norte, Av.\ Odilon Gomes de Lima  1722, 59078-400 Natal, Brazil}}

\begin{abstract}
We consider the Pseudo-Gap (PG) state of high-$T_{c}$ superconductors
in form of a composite order parameter fluctuating between 2${\bf {p}}_{\text{F}}$-charge
ordering and superconducting (SC) pairing. In the limit of linear
dispersion and at the hotspots, both order parameters are related
by a SU(2) symmetry and the eight-hotspot model of Efetov \textit{et
al}.\ {[}Nat.\ Phys.\ \textbf{9}, 442 (2013){]} is recovered. In
the general case however, curvature terms of the dispersion will break
this symmetry and the degeneracy between both states is lifted. Taking
the full momentum dependence of the order parameter into account,
we measure the strength of this SU(2) symmetry breaking over the full
Brillouin zone. For realistic dispersion relations including curvature
we find generically that the SU(2) symmetry breaking is small and
robust to the fermiology and that the symmetric situation is restored
in the large paramagnon mass and coupling limit. Comparing the level
splitting for different materials we propose a scenario that could
account for the competition between the PG and the SC states in the
phase diagram of high-$T_{c}$ superconductors.
\end{abstract}
\maketitle

\section{Introduction}

Reflecting our rather poor understanding of the physics of cuprate
superconductors, two kinds of theories are still debating whether
the final solution for this problem will be a ``bottom-up'' approach
based on a strong coupling theory \cite{Lee06,Gull:2013hh,Sorella02}
or rather a ``top-down'' approach, where symmetries and proximity
to a Quantum Critical Point (QCP) plays a dominant role \cite{Norman03,abanov03,sfbook,note}.
The recently proposed Eight Hot Spots (EHS) model is a promising ``top-down''
approach to cuprate superconductors \cite{Metlitski10b,Efetov13}.
It reduces the Fermi surface to only eight points on the anti-ferromagnetic
(AF) zone boundary and taking long-range AF fluctuations between them
into account. When the dispersion is linearized at the hot spots,
one observes surprisingly that an SU(2) symmetry relating the $d$-wave
SC channel (Cooper pairing) to the $d$-wave bond order, or Quadrupolar
Density Wave (QDW), (charge channel) is present. Moreover, an imposant
pre-emptive instability (of order of 0.6\,J, where J is the AF energy
scale) in the form of a composite SU(2) order parameter emerges, that
has been identified as a good candidate for the pseudo-gap (PG) state
of those compounds \cite{Efetov13,Meier13,Einenkel14,Meier14}. Motivated
by an impressive set of new experimental results \cite{Hoffman02,DoironLeyraud07,Wise08,Sebastian12,Ghiringhelli12,Wu13,LeBoeuf13,LeTacon14,Hayward14,Tabis14},
this theory points out to the emerging idea that charge order is most
certainly a key player in the physics of cuprate superconductors,
in addition to AF order, $d$-wave SC state and the Mott insulator
phase. Angle-resolved photoemission spectroscopy (ARPES) experiments
confirm as well the presence of modulations in the SC state for underdoped
Bi2201 \cite{He11,Vishik12} and has been interpreted either in terms
of charge order or Pairing Density Wave (PDW) inside the PG phase
\cite{Wang14,Agterberg:2014wf,Lee:2014ka,Fradkin:2014wk}. Finally,
we like to mention the recent interpretation of Raman resonances by
a collective SU(2) mode which is the first experimental result that
supports the idea of a composite PG \cite{Montiel15a}.

\begin{figure}
\centering \includegraphics[scale=0.5]{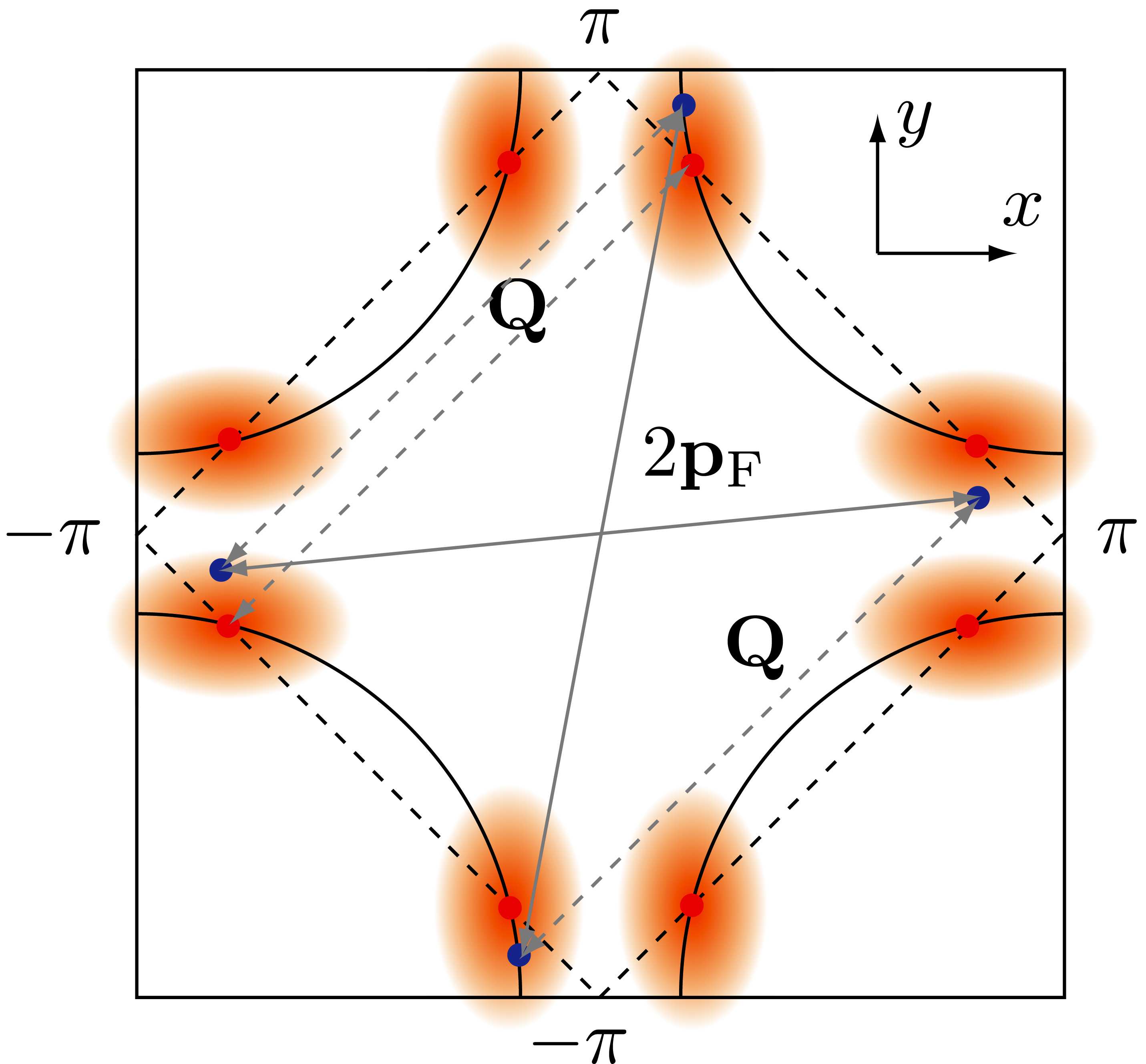} \vspace{-2.5ex}
 \caption{\label{fig1} (Color online) Schematic Fermi surface of hole-doped
superconductors in the first Brillouin zone of a square lattice. The
order parameters spatially extent over hot regions, that are centered
around the hotspots positions and which are coupled by the $2{\bf {p}}_{\text{F}}$
and the AFM coupling ${\bf {Q}}$ to opposed regions. }
\end{figure}

The idea of an emerging SU(2) symmetry belongs to a wide class of
theories which explain the PG phase of the cuprates through the notion
of degenerate symmetry states between the $d$-wave SC order and another
partner. Maybe the most famous attempt to such a unification has been
the SO(5) theory which relates the $d$-wave SC state to the AF order
\cite{Demler:2004tu}. Not less famous is the SU(2) symmetry which
relates the SC $d$-wave order to the $\pi$-flux phase of orbital
currents \cite{Lee06}. In both cases, the key question was to argue
that the energy splitting between the two orders was small enough
so that thermal effects would restore the symmetry above $T_{c}$
and below $T^{*}$, which are the SC and the PG critical temperatures.
The same question holds here for the SU(2) symmetry relating $d$-wave
and QDW order. Although the EHS model in its linearized version verifies
the symmetry exactly, it is not clear if a more realistic Fermi surface,
including curvature and the whole band structure, will induce small
or large energy splitting. Moreover, the EHS model relies on long-range
AF fluctuations which mediate the interactions, but the experimental
observation points out to short-range AF correlations which potentially
will gap a whole part of the Fermi surface, as depicted in Fig.\ \ref{fig1}.
A theory for ``hot regions'' instead of ``hot spots'' is thus
needed.

In this paper we address carefully all these issue by evaluating the
SU(2) splitting on realistic Fermi surfaces, for the two distinct
components of the composite order parameter: the $d$-wave $\chi$-field
in the charge sector (which forms the QDW order in the EHS model)
and the $d$-wave $\Delta$-field describing the SC pairing sector.
We find that the splitting of the SU(2) symmetry increases with the
mass of the paramagnons, but decreases with the strength of the coupling
constant between AF fluctuations and conduction electrons. This opens
a wide regime of parameters where the splitting is minimal -- of the
order of a few percents -- and where the SU(2) symmetry is expected
to be recovered through thermal effects in a regime of temperatures
$T_{c}<T<T^{*}$. Above the PG temperature $T^{*}$ all traces of
the short-range charge and SC field have disappeared. Of course the
SU(2) symmetry holds in the whole temperature range between $T_{c}$
and $T^{*}$, hence subleading charge instabilities which occur below
$T^{*}$ do have their SU(2) partners in the form of Pairing Density
Waves (PDW) \cite{Pepin14,Wang15a,*Wang15b}. We also study the effects
of the Fermi surface shape in breaking the SU(2) symmetry -- which
is only preserved in the EHS model with a linearized dispersion. We
find that the splitting is small away from the points where $\chi$
and $\Delta$ are maximal. In the physical situation of a large paramagnon
mass and also strong coupling, the maximum of $\chi$ moves towards
the zone edge leading to bond order parallel to the $x$-$y$ axes.
All these findings point to the realization that, while being a secondary
instability to AF ordering, charge order is a key player in the physics
of the PG phase of the cuprates.

\begin{figure}
\centering \includegraphics[scale=0.8]{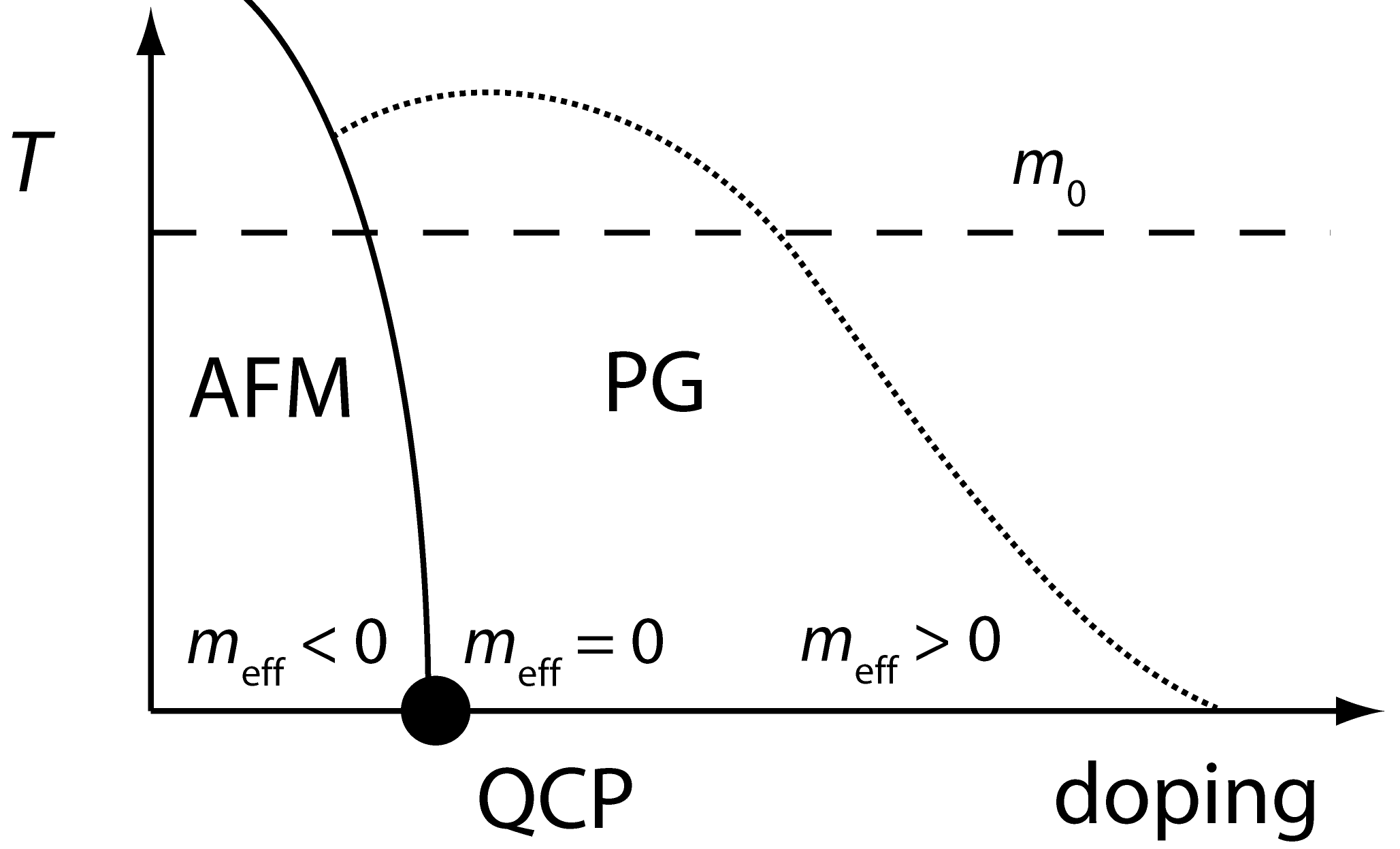} \vspace{-2.5ex}
 \caption{\label{fig2} Schematic phase diagram of hole-doped cuprate superconductors.
The effective mass of the paramagnon propagator $m_{\text{eff}}$
serves as as measure for the distance to the quantum critical point
QCP. In the SC phase $m_{\text{eff}}$ is positive and vanishes at
the QCP to become negative in the AFM phase. The bare mass $m_{\text{bare}}$
is defined far away from criticality at some higher temperature, indicated
by the dashed line. }
\end{figure}

\begin{figure}
\centering a)\includegraphics[width=75mm]{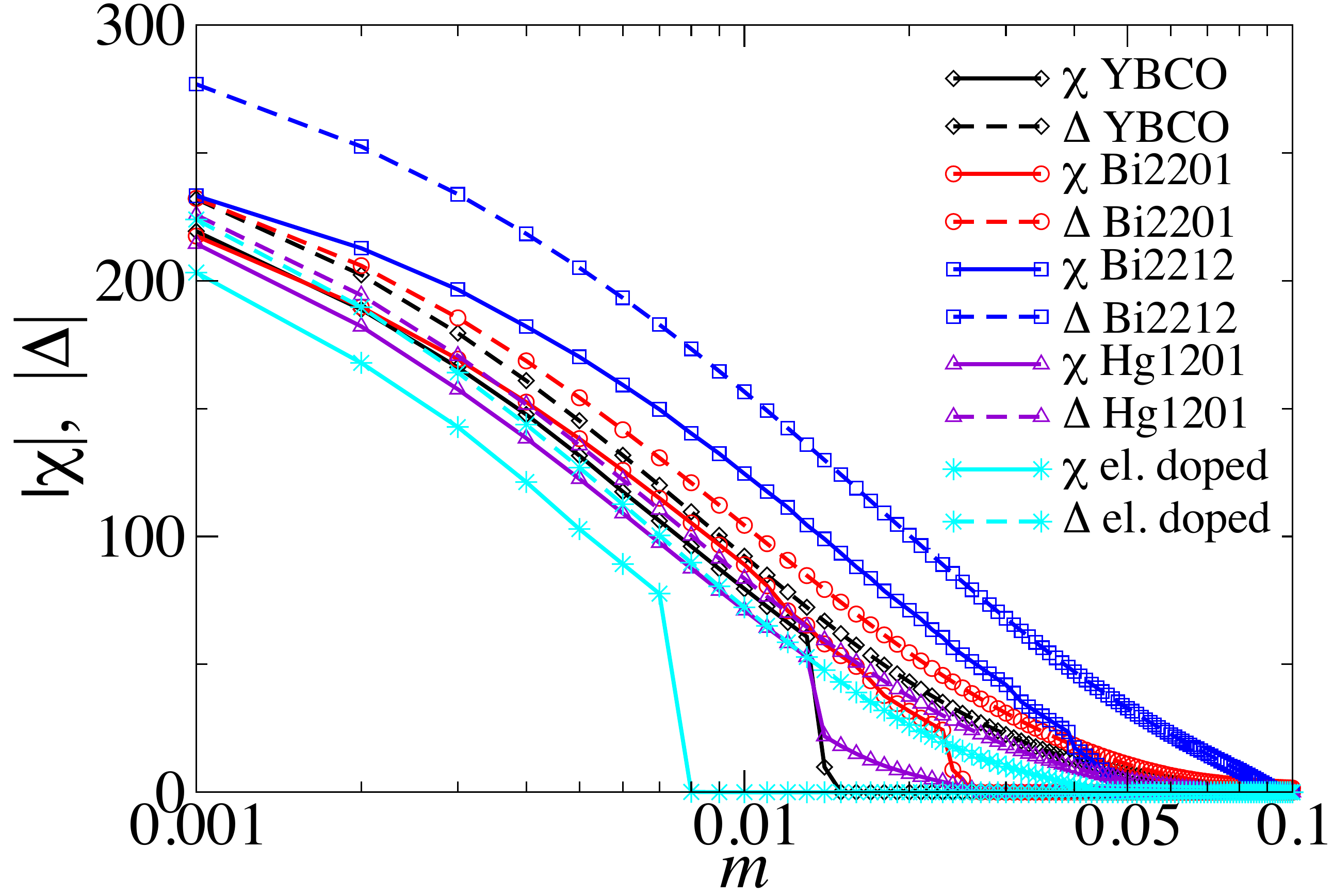} \vspace{0ex}
 b)\includegraphics[width=75mm]{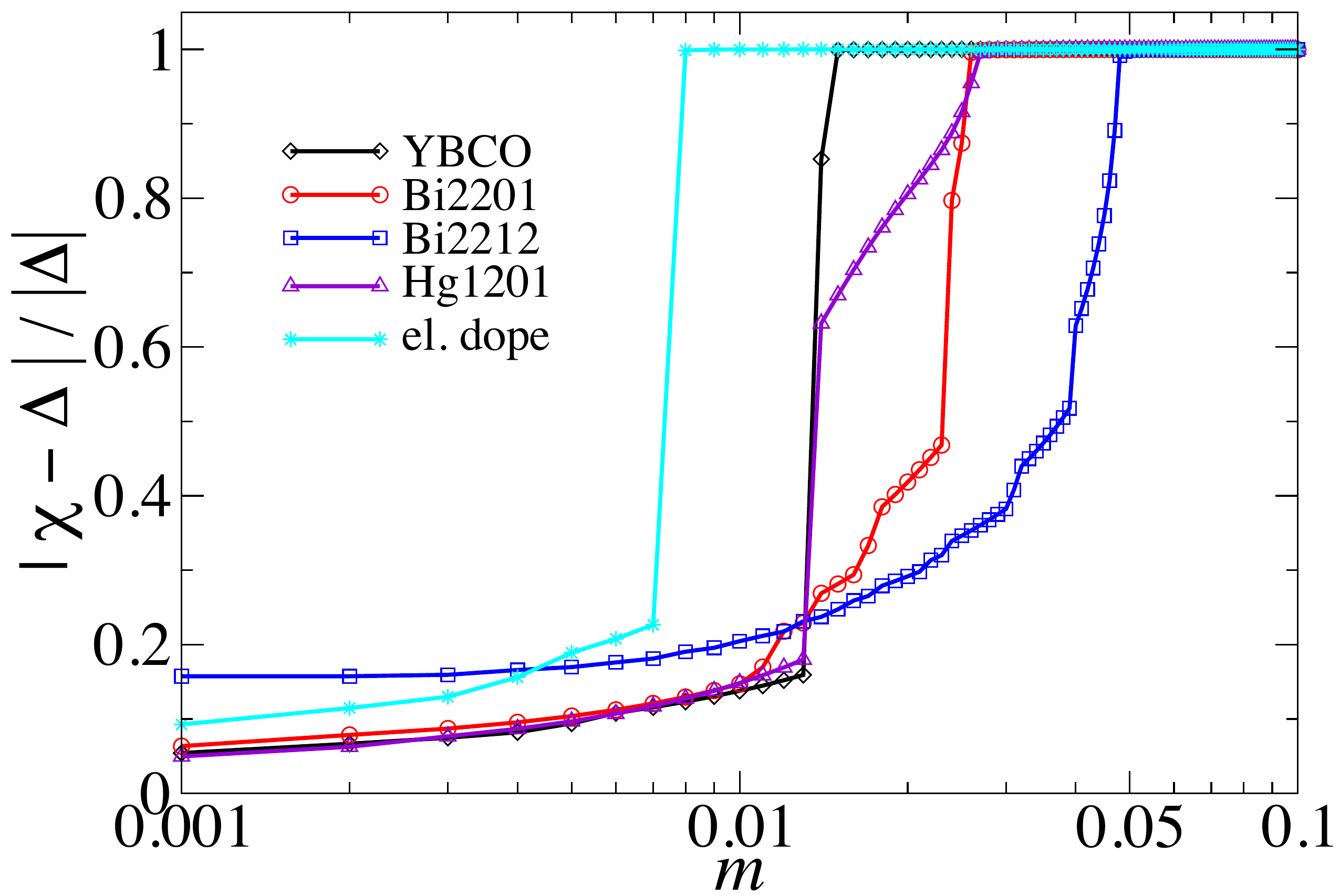} \vspace{-1ex}
 \caption{\label{fig3} (Color online) Panel a): Variation of the maximum value
of the gap functions $|\chi|$ and $|\Delta|$ as a function of the
mass $m$. Note that in all materials the 2${\bf {p}}_{\text{F}}$
pairing in terms of $|\chi|$ vanishes abruptly, whereas the SC pairing
in terms of $|\Delta|$ approaches zero asymptotically when the paramagnon
mass $m$ is increased. Panel b): Variation of the level splitting
as a function of the mass $m$. In both panels $\lambda=44$. }
\end{figure}

\begin{figure}
\centering a)\includegraphics[width=75mm]{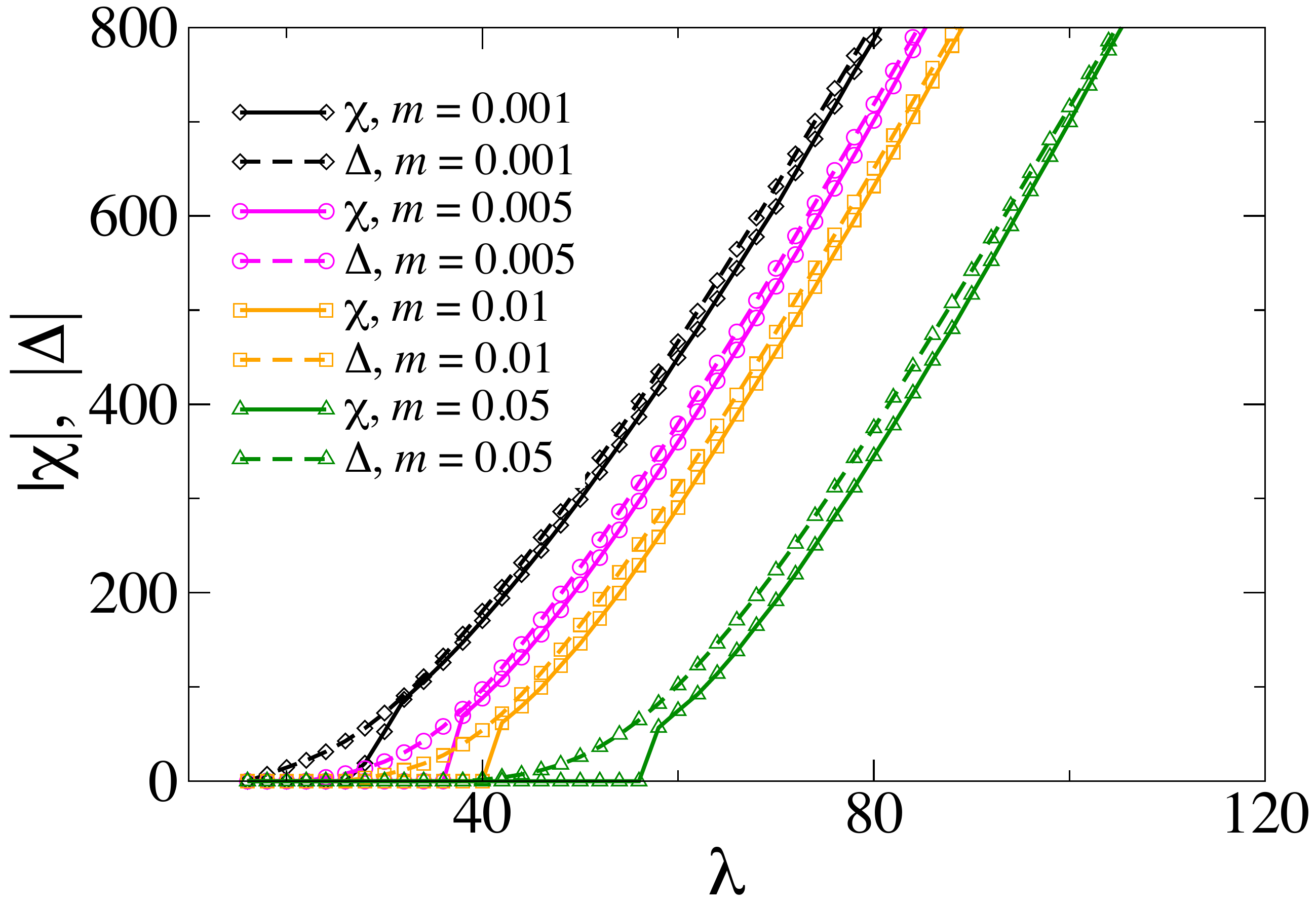} 
b)\includegraphics[width=75mm]{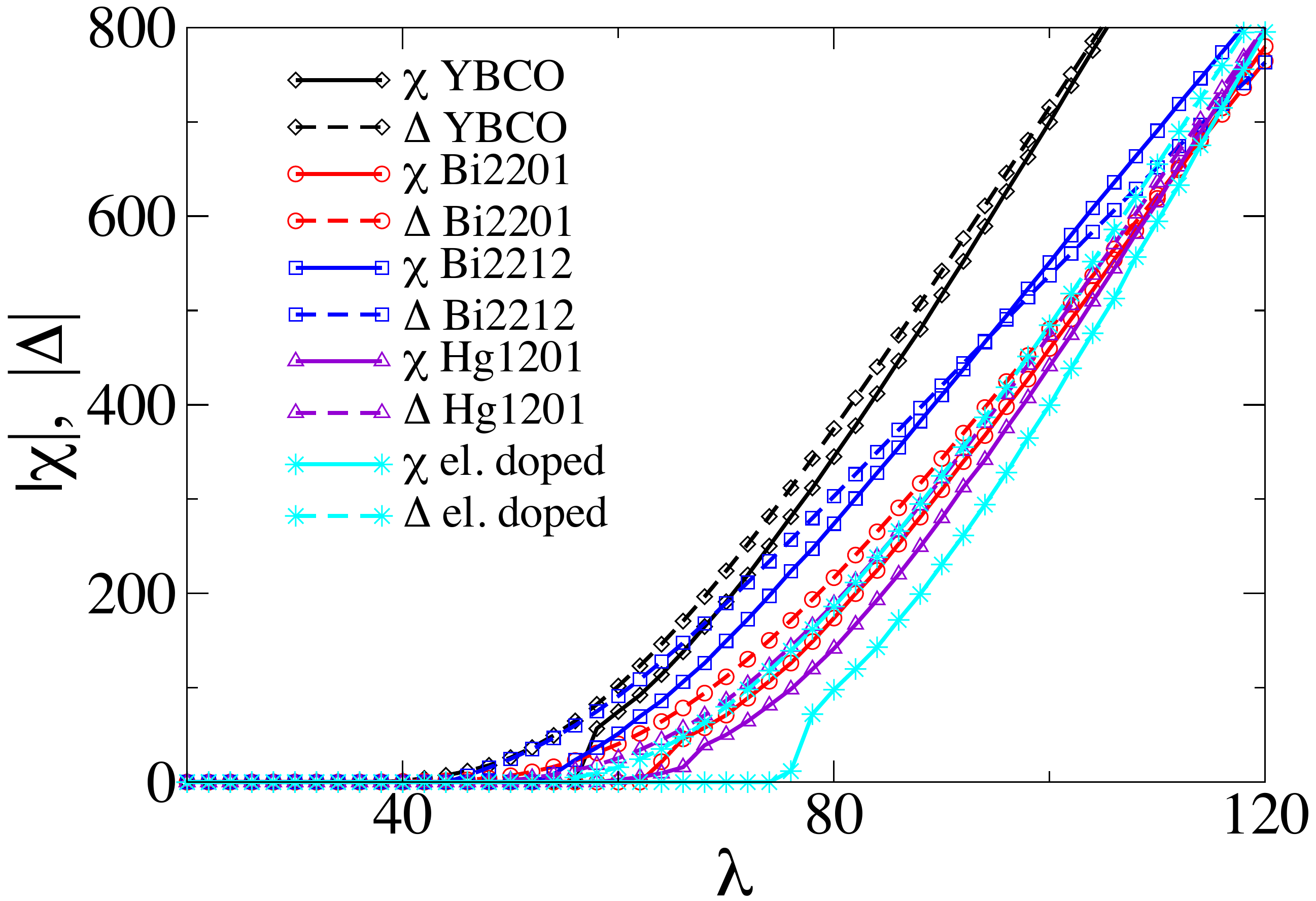}
\caption{\label{fig4} (Color online) Panel a): Variation of the maximum value
of the gap functions $|\chi|$ and $|\Delta|$ in YBCO as a function
of the coupling $\lambda$ for different masses $m$. Panel b): Variation
of the maximum value of the gap functions $|\chi|$ and $|\Delta|$
for different compounds as a function of the coupling $\lambda$ for
the mass $m=0.5$. }
\end{figure}

\section{Model}

We start from the spin-fermion model \cite{Metlitski10b,Efetov13,Pepin14}
with Lagrangian $L=L_{\psi}+L_{\phi}$, where \begin{subequations}
\begin{align}
L_{\psi} & =\psi^{*}\left(\partial_{\tau}+\epsilon_{k}+\lambda\phi\sigma\right)\psi\ ,\\
L_{\phi} & =\frac{1}{2}\phi D^{-1}\phi+\frac{u}{2}\left(\phi^{2}\right)^{2}\ .
\end{align}
\label{eq2} \end{subequations} The fermionic field $\psi$ describes
the electrons which are coupled via $L_{\phi}$ to spin waves described
by the bosonic field $\phi$. The effective spin-wave propagator is
$D_{q}^{-1}=\gamma|\omega|+{|{\bf {q}|}}^{2}+m$ where $m$ is the
paramagnon mass which vanishes at the QCP and $\gamma$ a phenomenological
coupling constant, which we estimate from its form in the EHS model
\cite{Efetov13} to be of the order $10^{-5}$. For notational reasons
we also write $q\equiv(i\omega,{\bf {q})}$. Neglecting the spinwave
interaction ($u=0$) one can formally integrate out the bosonic degrees
of freedom. In the spin boson model, this generates an effective
spin-spin interaction of the form 
\begin{equation}
S_{\text{int}}=-\sum_{q}\bar{J}_{q}\vec{S}_{q}\vec{S}_{-q}.
\end{equation}
It is convenient to use a fermionic representation of the spin operator
$\vec{S}$ and consider in the following only the paramagnetically
ordered phase in $z$-direction, so that $\bar{J}=3/2J$. The partition
function then writes $Z=\int\mathcal{D}[\Psi]\exp(-S_{0}-S_{int})$
with \begin{subequations} 
\begin{align}
S_{0} & =\sum_{k,\sigma}\Psi_{k}^{\dagger}G_{0,k}^{-1}\Psi_{k},\\
S_{int} & =-\sum_{k,k',q,\sigma}J_{q}\psi_{k,\sigma}^{\dagger}\psi_{k+Q+q,\bar{\sigma}}\psi_{k',\bar{\sigma}}^{\dagger}\psi_{k'-q-Q,\sigma}.\label{eq:s1}
\end{align}
\end{subequations} where the bare propagator is 
\begin{equation}
\hat{G_{0}}_{k}^{-1}=\diag(i\omega-\epsilon_{{\bf {k}}},i\omega+\epsilon_{-{\bf {k}-{\bf {p}}}},i\omega-\epsilon_{{\bf {k}+{\bf {p}}}},i\omega+\epsilon_{-{\bf {k}}}),\label{eq:g0m1}
\end{equation}
and the spinor field $\Psi_{k}=(\psi_{k,\sigma},\psi_{-k-p,\bar{\sigma}}^{\dagger},\psi_{k+p,\sigma},\psi_{-k,\bar{\sigma}}^{\dagger})^{T}$.
Furthermore, $J_{q}^{-1}=4D_{q}^{-1}/3\lambda^{2}$, $\sigma\in\{\uparrow,\downarrow\}$
labels the spin, ${\bf {Q}=(\pi,\pi)}^{T}$ is the AFM ordering vector
and ${\bf {p}}$ stands for the 2${\bf {p}}_{\text{F}}$ vector, as
depicted in Fig.\ \ref{fig1}. Note that the chemical potential $\mu$
is implicitly subtracted from the dispersion $\epsilon_{{\bf {k}}}$.
We select the SC and the 2${\bf {p}}_{\text{F}}$ channel by introducing
the two order parameters 
\begin{equation}
\Delta_{k}=\langle\psi_{k,\sigma}^{\dagger}\psi_{-k,\bar{\sigma}}^{\dagger}\rangle,\qquad\chi_{k}=\langle\psi_{k,\sigma}^{\dagger}\psi_{k+p,\sigma}\rangle.
\end{equation}
The interaction $S_{1}$ is now decoupled by means of a Hubbard-Stratonovich
transformation. The partition function becomes (up to a normalization
factor) 
\begin{equation}
Z=\int\mathcal{D}[\Psi]\mathcal{D}[\Delta,\chi]\exp[-S_{0}-S_{1,eff}].\label{eq:partFunc2}
\end{equation}
The effective interaction is 
\begin{align}
S_{1,eff}=\sum_{k,q,\sigma}\Bigl[J_{q}^{-1}\chi_{k}^{\dagger}\chi_{\bar{k}+q}+J_{q}^{-1}\Delta_{k}^{\dagger}\Delta_{\bar{k}+q}\Bigr]-\sum_{k,\sigma}\Psi_{k}^{\dagger}\hat{M}_{k}\Psi_{k},
\end{align}
with $\bar{{\bf {k}}}={\bf {k}+{\bf {Q}}}$ and the matrix $\hat{M}$
is 
\begin{equation}
\hat{M}_{k}=\left(\begin{matrix} & \hat{m}_{k}\\
\hat{m}_{k}^{\dagger}
\end{matrix}\right),\qquad\hat{m}_{k}=\left(\begin{matrix}-\chi_{k} & -\Delta_{k}\\
-\Delta_{k+p}^{\dagger} & \chi_{-k}
\end{matrix}\right).
\end{equation}
The fermions in Eq.\ (\ref{eq:partFunc2}) can now be integrated
out so that the partition function becomes 
\begin{equation}
Z=\int\mathcal{D}[\hat{M}]\exp\Bigl[-\frac{1}{4}\sum_{k,q}\tr\, J_{q}^{-1}\hat{M}_{\bar{k}+q}\hat{M}_{k}+\frac{1}{2}\sum_{k}\tr\log\hat{G}_{k}^{-1}\Bigr],
\end{equation}
with $\hat{G}^{-1}=\hat{G}_{0}^{-1}-\hat{M}$. After functional differentiation
of the free energy $F=-T\ln Z$ with respect to $\hat{M}_{k}$ we
obtain the MF equations in matrix form 
\begin{equation}
\hat{M}_{k}=\sum_{k'}J_{\bar{k}-k'}\hat{G}_{k'}.
\end{equation}
The matrix equation can now be projected onto the different components.
We will consider here the case of two competing order parameters which
can not be non-zero at the same point in ${\bf {k}}$ space. Therefore,
we consider the equation for $\Delta$ with $\chi=0$ and vice versa.
The gap equations follow as \begin{subequations} 
\begin{align}
\Delta_{k} & =T\sum_{\omega',{\bf {k}'}}J_{\bar{k}-k'}\frac{\Delta_{k'}}{\Delta_{k'}^{2}+\epsilon_{{\bf {k}'}}^{2}+{\omega'}^{2}},\\
\chi_{k} & =-\Re T\sum_{\omega',{\bf {k}'}}J_{\bar{k}-k'}\frac{\chi_{k'}}{(i\omega'-\epsilon_{{\bf {k}'}})(i\omega'-\epsilon_{{\bf {k}'+{\bf {p}}}})-\chi_{k'}^{2}}.
\end{align}
\label{eq:gapeq} \end{subequations} To solve these equations numerically,
$\epsilon_{{\bf {k}}}$ is parametrized in tight-binding approximation
with the following parameters: YBCO \cite{Norman07} (parameter set
tb2), Bi2201 \cite{He14}, Bi2212 \cite{Fujita14} and Hg1201 \cite{Das12}
and for electron doped cuprates \cite{Senechal04}. The momentum sums
in Eq.\ (\ref{eq:gapeq}) are then carried out by discretizing the
${\bf {k}}$-space by rectangular and equidistant grids. To keep the
numerical computations tractable we neglect the frequency dependence
of $\chi$ and $\Delta$ %
\footnote{We have checked that this approximation is reliable to calculate the
spacial distribution of the gap functions.%
}. The Matsubara sums are then carried out exactly in the limit $T\rightarrow0$
and the momentum sums are performed over $200\times200$ points and
over one Brillouin Zone (BZ). Moreover, note that the 2${\bf {p}}_{\text{F}}$
vector which connects two opposed FS points at $\pm{\bf {p}}_{\text{F}}$
depends on the external momentum ${\bf {k}}$ in Eq.\ (\ref{eq:gapeq})
and is only properly defined on the FS. Since we expect that the main
contribution to the momentum sum in Eq.\ (\ref{eq:gapeq}) comes
however from the hot-spot region, we make the approximation to take
the 2${\bf {p}}_{\text{F}}$ vector constant and take the 2${\bf {p}}_{\text{F}}$
from the hotspot for arbitrary points in the first BZ, as depicted
in Fig.\ \ref{fig1}. Throughout this article, we use $10\%$ hole
filling (respectively $10\%$ electron filling in the electron doped
case) and the bandgap is 10$^{4}$\,K. To evaluate the strength of
the SU(2) symmetry, we study the level splitting $|\chi-\Delta|/\Delta$.
This parameter afford the study of the relative amplitude between
the QDW and SC order parameter, $\chi$ and $\Delta$. It vanishes
for a perfect SU(2) symmetry and becomes closer to one for a complete
SU(2) symmetry breaking. From the Lagragian we find that $\phi\lambda$
has dimension of an energy and $\phi\sim m^{-1/2}$. To estimate the
coupling strength, we evaluate an effective energy via $E_{\text{eff}}=\phi\lambda\sim m^{-1/2}\lambda$
and away from criticality $E_{\text{0}}\sim m_{0}^{-1/2}\lambda$
where we take $m_{0}=1$ as bare mass, compare Fig.\ \ref{fig2}.

\begin{widetext}

  \begin{figure}
   \centering
 \begin{minipage}{4.2cm}
\includegraphics[width=45mm]{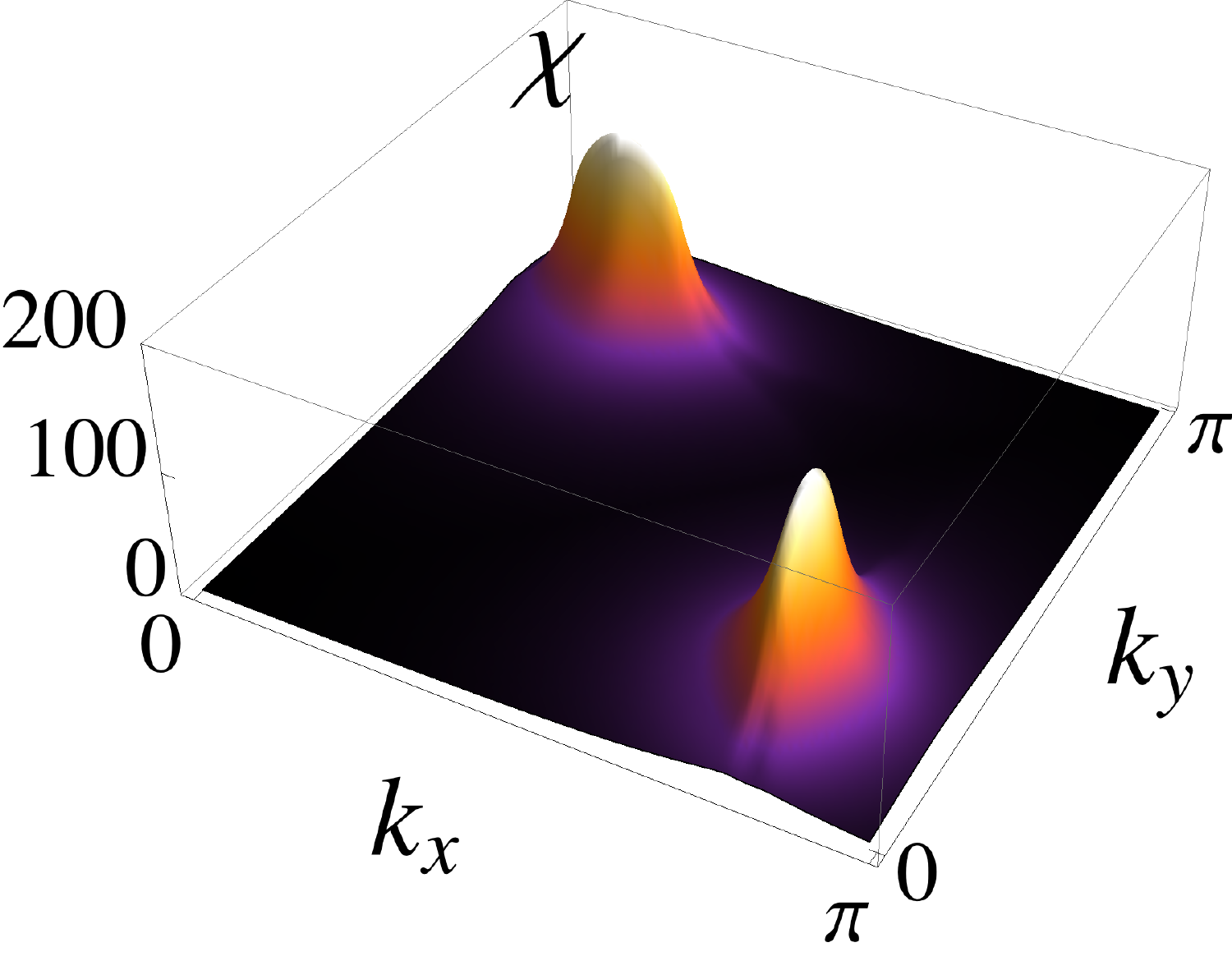}
\end{minipage}
 \begin{minipage}{4.2cm}
\includegraphics[width=45mm]{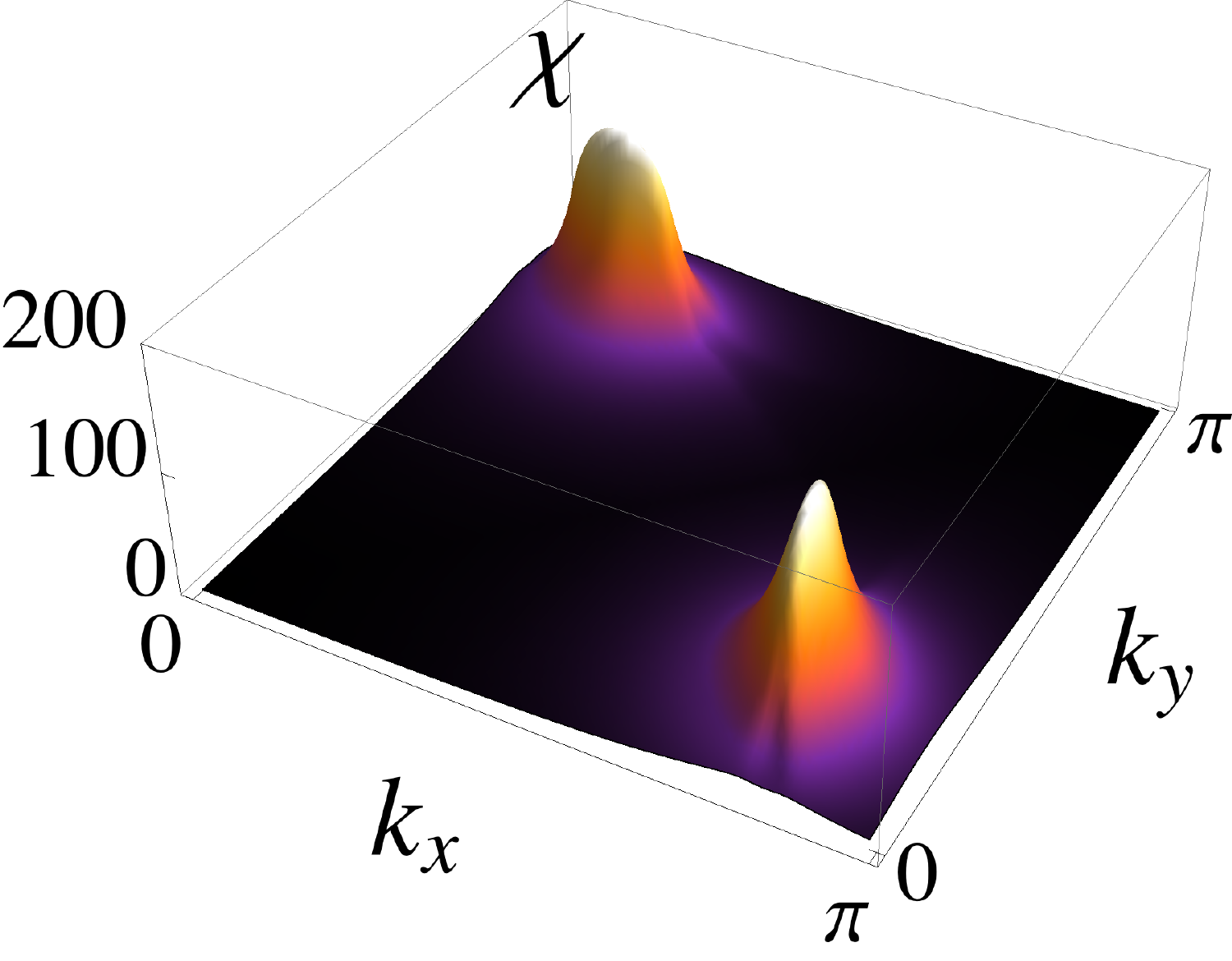}
\end{minipage}
\begin{minipage}{4.2cm}
\includegraphics[width=45mm]{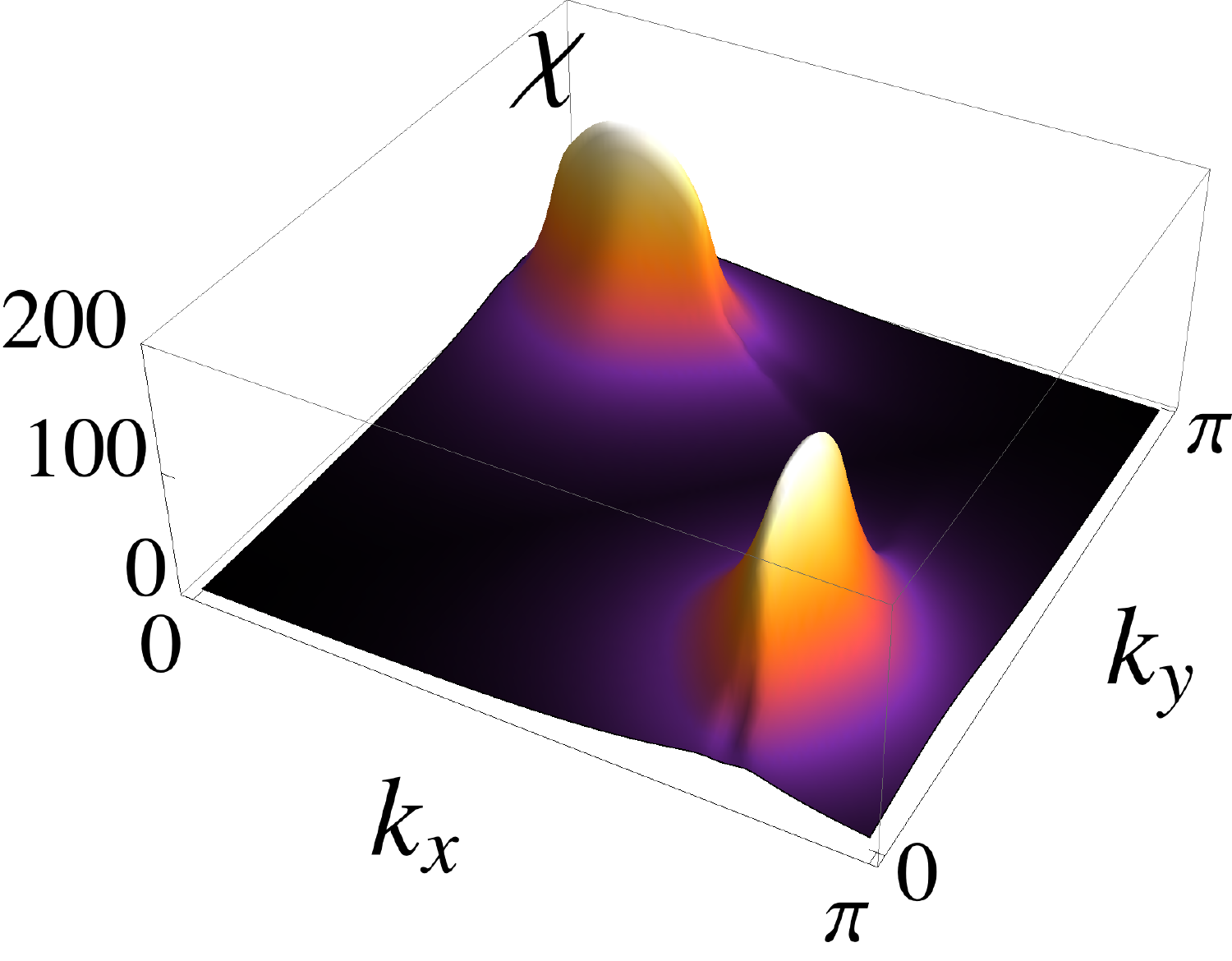}
\end{minipage}
\begin{minipage}{4.2cm}
\includegraphics[width=45mm]{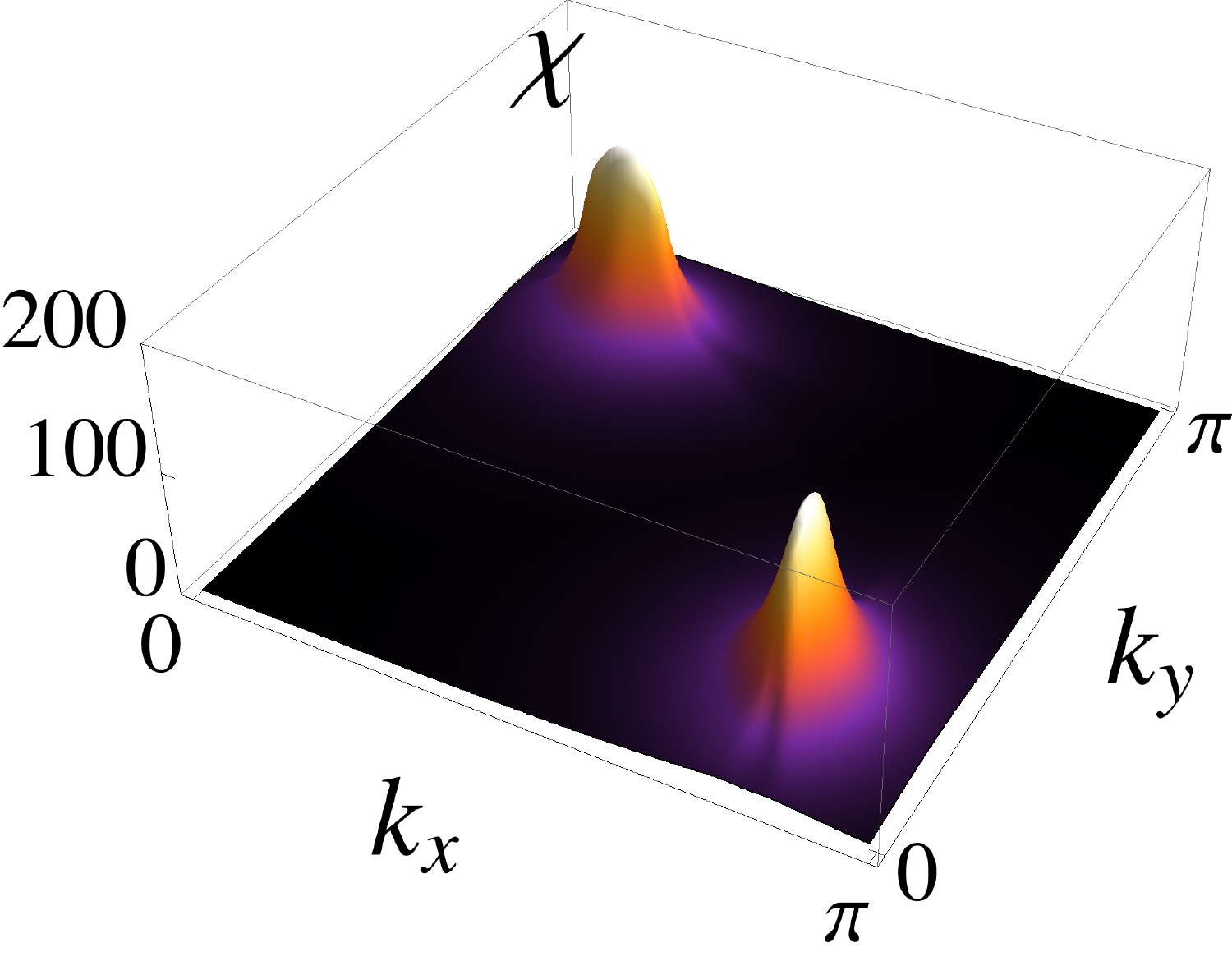}
\end{minipage}
   \begin{minipage}{4.2cm}
\includegraphics[width=45mm]{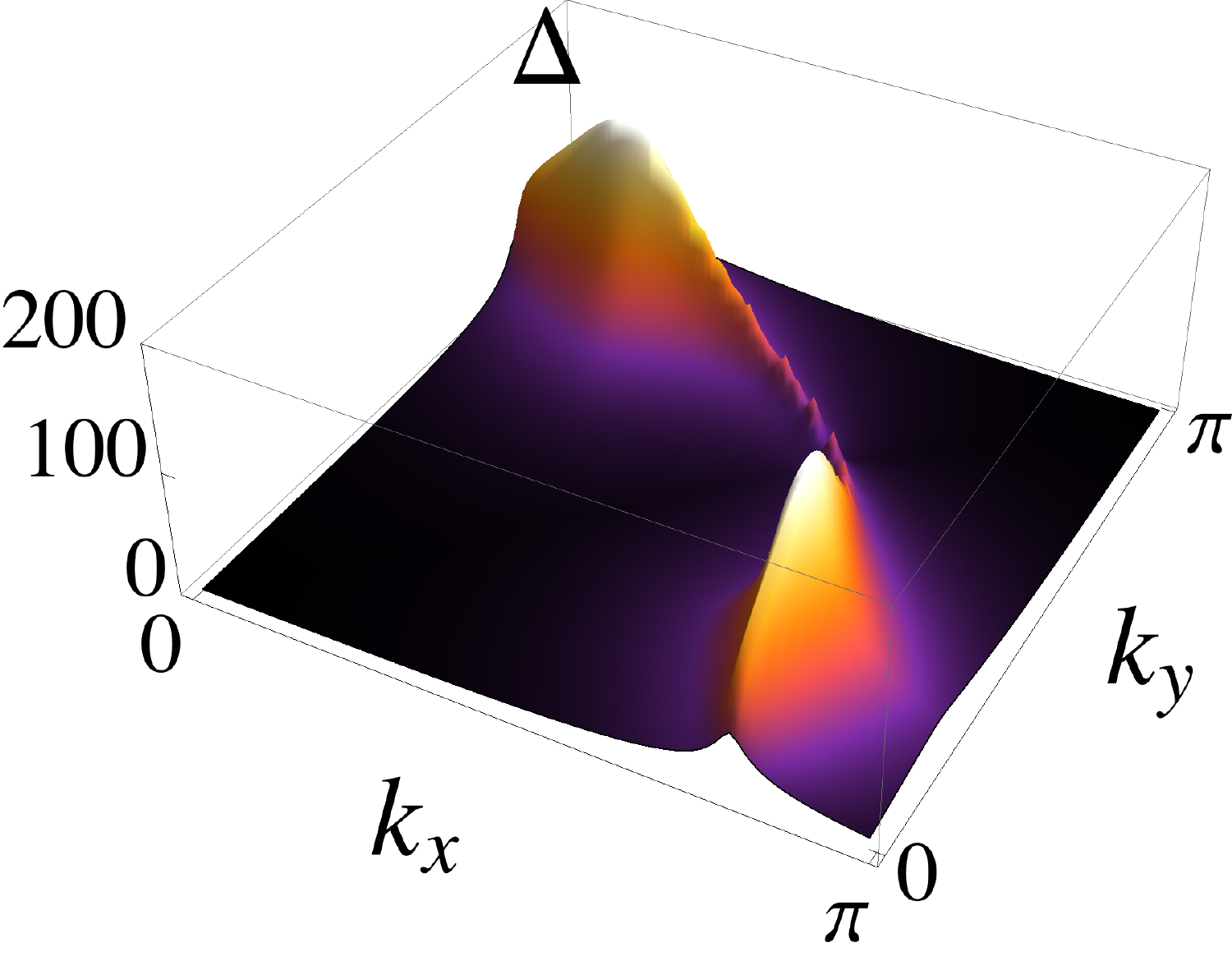}
\end{minipage}
 \begin{minipage}{4.2cm}
\includegraphics[width=45mm]{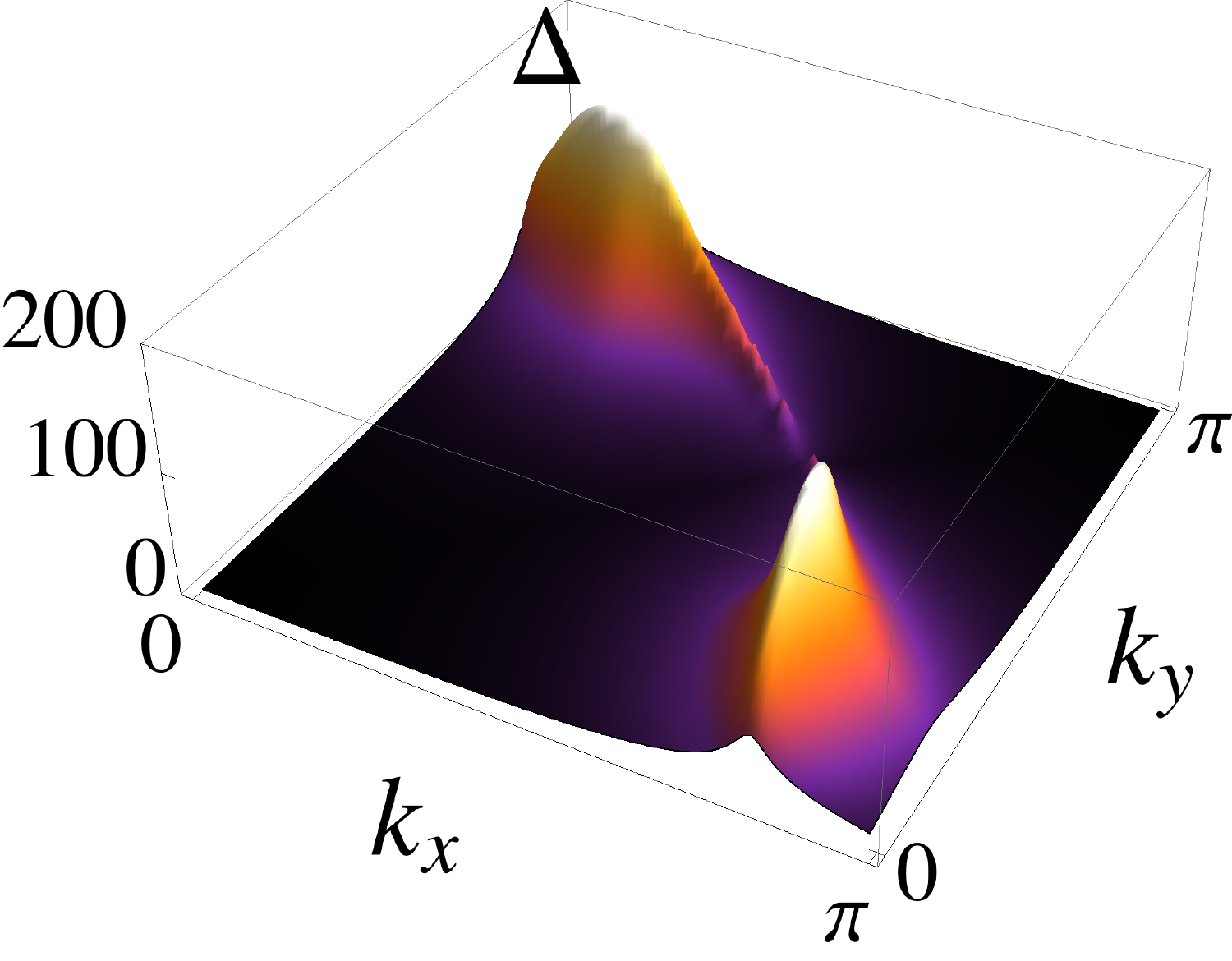}
\end{minipage}
\begin{minipage}{4.2cm}
\includegraphics[width=45mm]{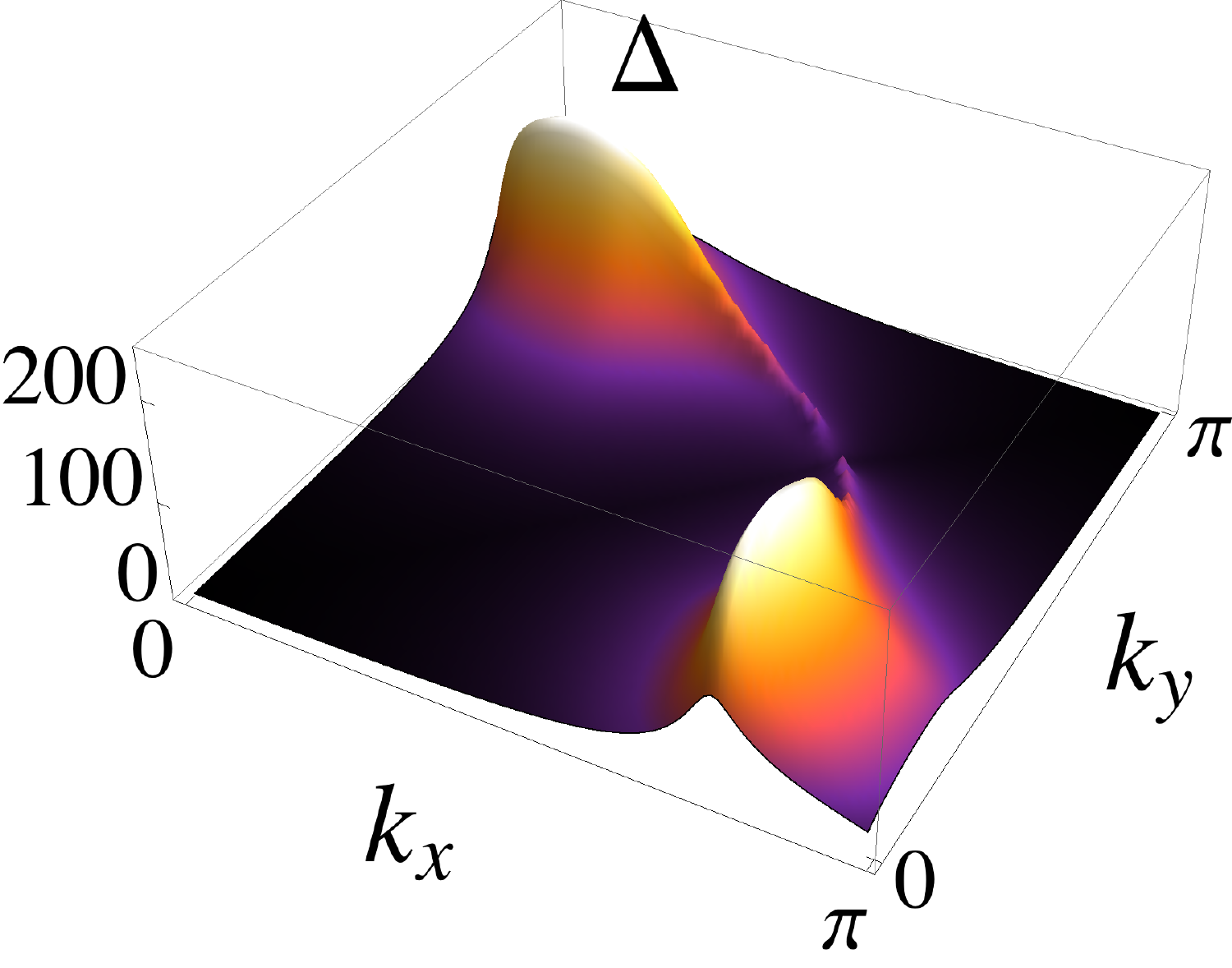}
\end{minipage}
\begin{minipage}{4.2cm}
\includegraphics[width=45mm]{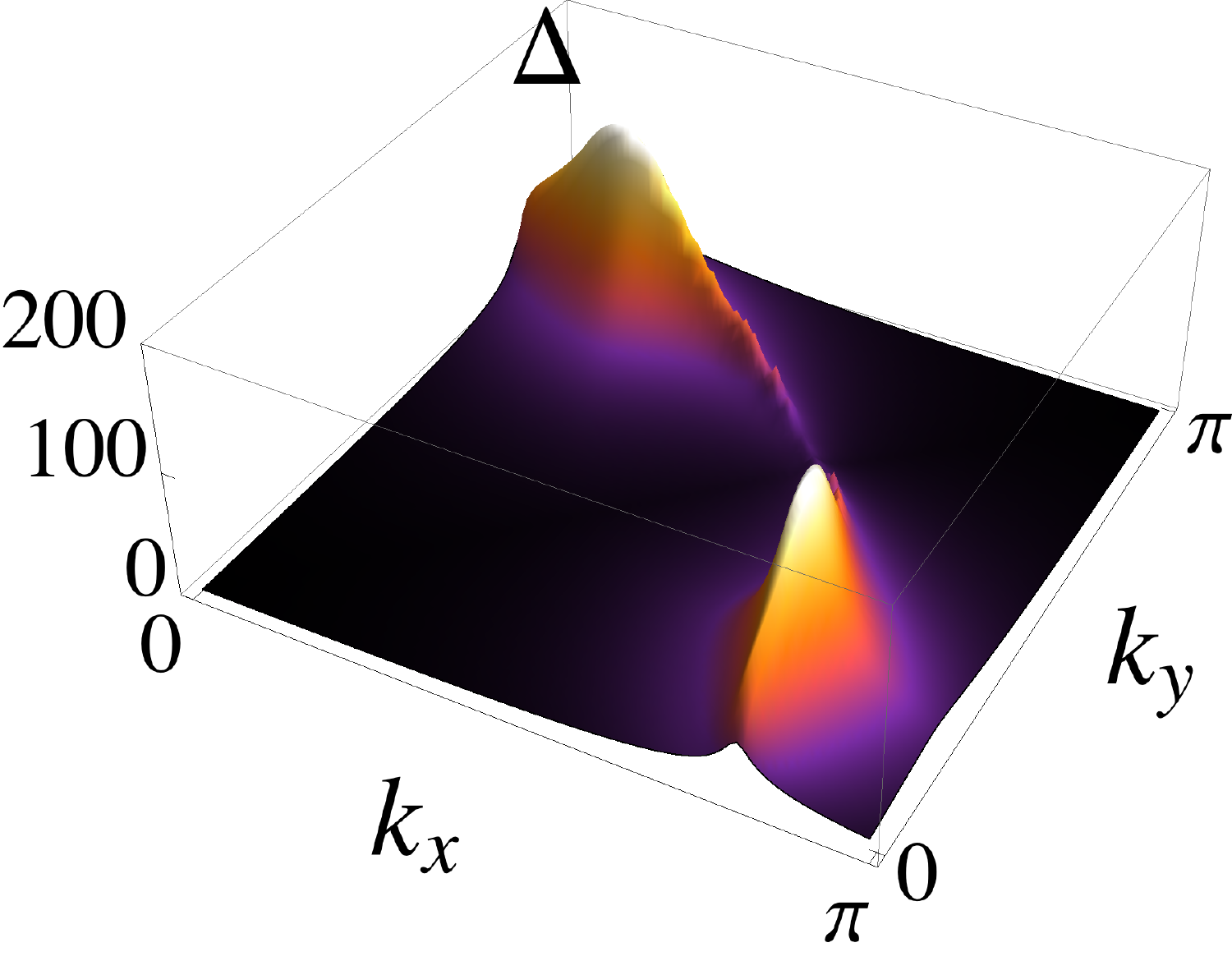}
\end{minipage}
   \begin{minipage}{4.2cm}
\includegraphics[width=36mm]{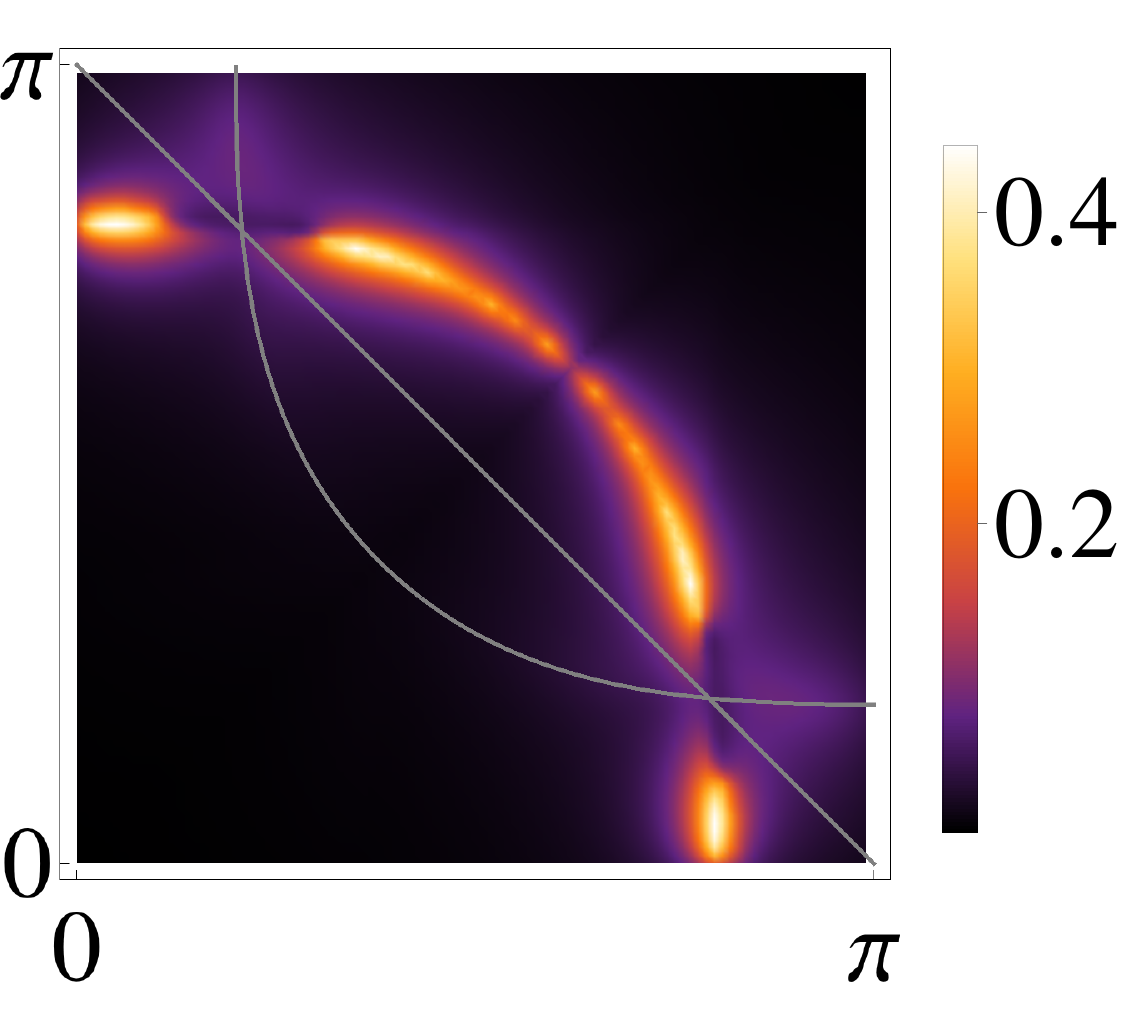}
\end{minipage}
 \begin{minipage}{4.2cm}
\includegraphics[width=36mm]{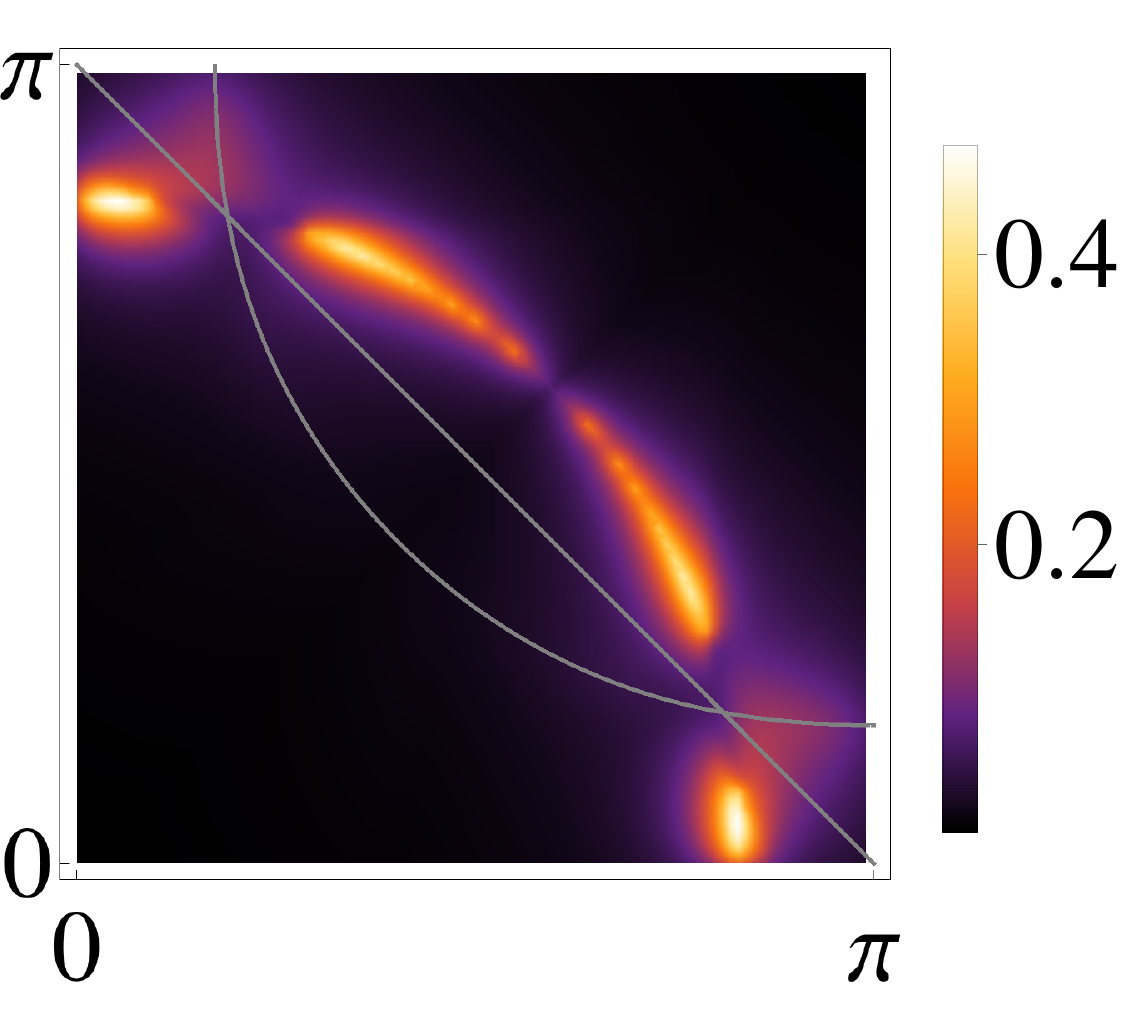}
\end{minipage}
\begin{minipage}{4.2cm}
\includegraphics[width=36mm]{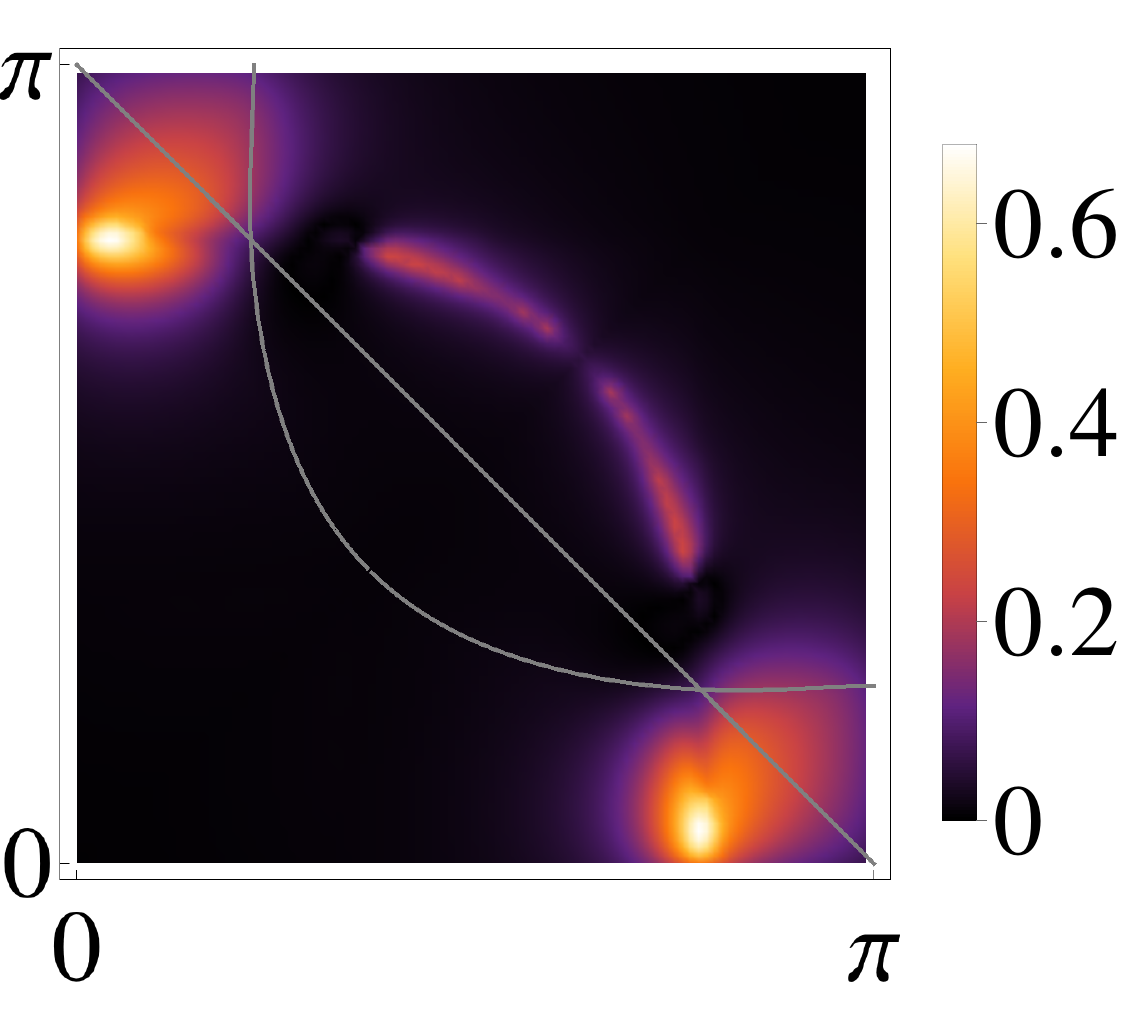}
\end{minipage}
\begin{minipage}{4.2cm}
\includegraphics[width=36mm]{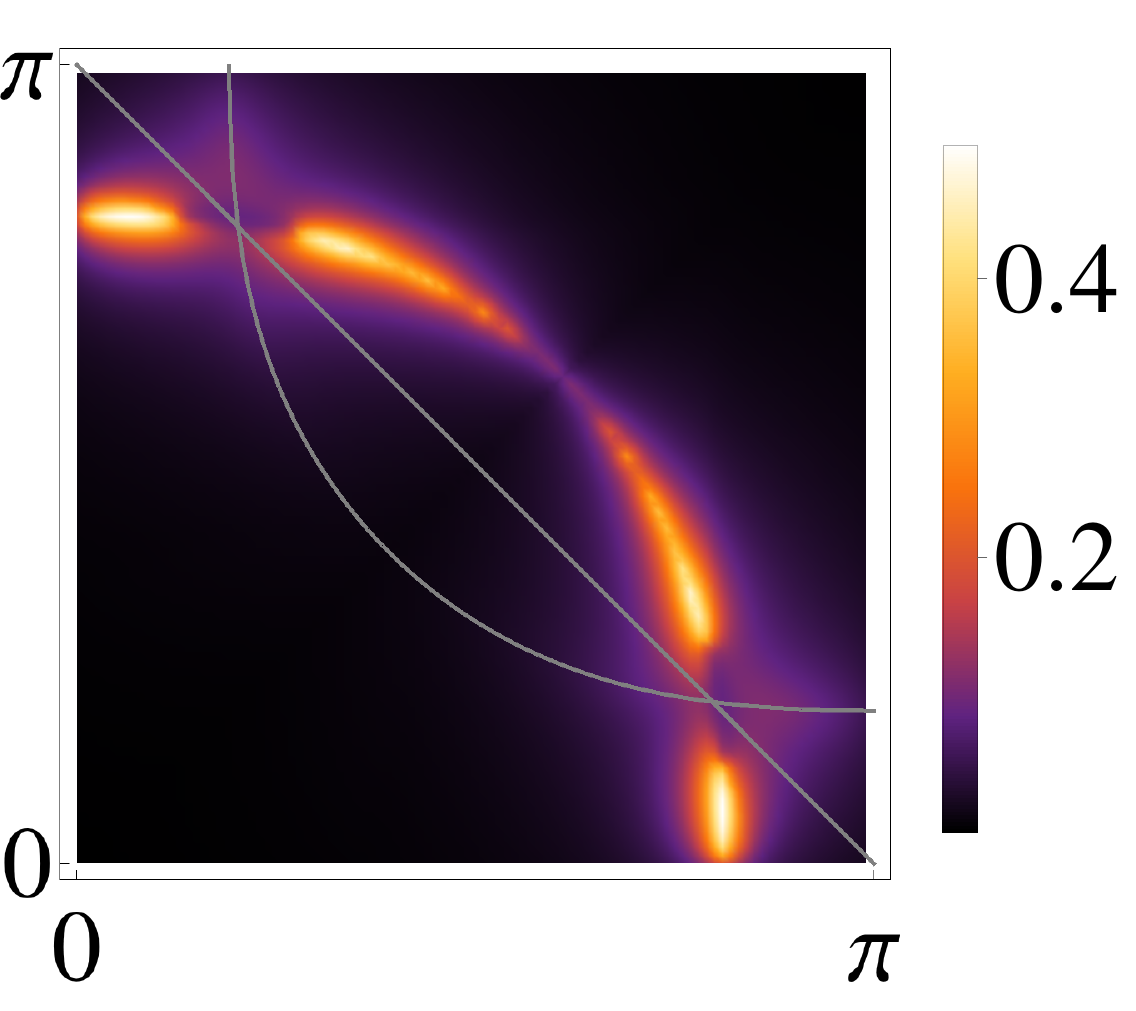}
\end{minipage}
  \vspace{-2mm}
 \caption{\label{fig5} (Color online) Gap functions $|\chi|$, $|\Delta|$
and $|\chi-\Delta|/|\Delta_{\text{max}}|$ (from up to down) for different
materials in the first BZ. The compounds are (from left to right):
YBCO, Bi2201, Bi2212 and Hg1201. The figures correspond to the small
mass and small coupling limit ($\lambda=40$ and mass $m=10^{-3}$;
note that the bare coupling is small w.r.t.\ the renormalized one
$E_{\text{0}}\simeq40\,$K and $E_{\text{eff}}=1265\,$K) that is
most unfavorable for the SU(2) symmetry.}
\end{figure}

\end{widetext}

\section{Results and Discussion}

\subsection{Mass and coupling dependence on level splitting}

In order to test the effect of the curvature on the level degeneracy,
we have plotted in Fig.\ \ref{fig3}a) the variation of the maxima
of $\chi$ and $\Delta$ with the paramagnon mass $m$ for a fixed
value of the coupling constant $\lambda$. We observe a similarity
between the various compounds that we have tested. In a wide range
at low value of the mass, the SU(2) degeneracy between $\chi$ and
$\Delta$ is verified within a few percents. The existence of such
a regime is an indication that a PG driven by SU(2) symmetry is possible
in cuprate superconductors. As the mass is increased we progressively
lose the level degeneracy with the parameter $\chi$ abruptly dropping
down while the paring $\Delta$ is asymptotically going down to zero
when the mass increases. It is interesting to see that the
SU(2) symmetry is weak for the electron doped and Hg1201 compound
which experimentally show much weaker signs of charge order \cite{Tabis14,comin14a}.
We also find that the compound Bi2212 behaves sightly different
than the other compounds in Fig.\ \ref{fig3} and \ref{fig4}, although
it is not clear at the current stage where this deviation comes from.
In Fig.\ \ref{fig3}b) the level splitting is directly shown for
all the compounds and the two regimes, the one at low mass where the
SU(2) symmetry is obtained and the higher mass regime where $|\chi-\Delta|/\Delta$
becomes of order one are clearly seen. Within the non linear $\sigma$-model
associated to the present theory \cite{Efetov13}, the SU(2) regime
is a signature of the PG of the system, while the energy splitting
of the two levels is associated to the superconducting $T_{c}$. We
see that Fig.\ \ref{fig3}a) mimics the generic phase diagram of
the cuprates where the PG line $T^{*}$ abruptly plunges inside the
SC dome at some value of oxygen doping.

Although it is very encouraging to see that the SU(2) regime has a
non-zero probability to exist, one can wonder whether the paramagnon
mass in hole doped cuprate superconductors is small, since typically
the AF correlation length is of few lattice constants \cite{Hinkov04}.
The issue is addressed in Fig.\ \ref{fig4}a), where the values of
$\chi$ and $\Delta$ are shown for a fixed mass as a function of
the coupling constant $\lambda$. Here again a generic pattern emerges.
For small $\lambda$ the SU(2) symmetry is broken, but surprisingly,
above a certain threshold of $\lambda$, the SU(2) symmetry is almost
completely restored. As seen in Fig.\ \ref{fig4}a), the bigger the
mass is, the stronger the coupling constant needs to be for the symmetry
to be restored. Figure \ref{fig4}b) shows that this behavior
is quite general among the different compounds. Note however that
the electron -doped compound is less sensitive to the effect of increasing
the coupling constant, compared to the other hole-doped ones, for
which the SU(2) symmetry is restored for large enough $\lambda$.

\subsection{Spacial dependence of the splitting}

The SU(2) symmetry is not only broken due to the curvature of the
Fermi surface at the hot spots, but it is typically broken in the
BZ away from the eight hot spots. Figure \ref{fig5} shows the typical
shape of $\left|\chi\right|$ and $\left|\Delta\right|$ for four
compounds under investigation for the most unfavorable case for the
symmetry, that is for small values of the mass and coupling constant.
The level splitting is shown
as a density plot in the bottom. It is rather small almost everywhere
in the BZ and at the hotspot positions with maxima of the order of
20-40\% around the ``shadow'' Fermi surface. The main learning from
these plots is that the variations of the Fermi surface geometry gives
a rather small departing from the SU(2)-degeneracy for a various range
of compounds. In all cases, the SU(2) symmetry is well respected
at the hotspot positions.

In Fig.\ \ref{fig6} we place ourselves in the strong coupling and
strong mass regime and plot the variation of $\left|\chi\right|$
, $\left|\Delta\right|$ and $|\chi-\Delta|/\Delta$. The level splitting
is also shown in Fig.\ \ref{fig6} and found to be much smaller than
the previous case in Fig.\ \ref{fig5}, and has dropped to an order
of 5-10\% %
\footnote{The gap functions and level splitting of the other compounds are very
similar and therefore not shown in Fig.\ \ref{fig6}%
}. Interestingly, the typical shape of $\left|\chi\right|$ and $\left|\Delta\right|$
in the BZ has changed compared to Fig.\ \ref{fig5}, with maxima
now around the zone edge. This is the justification that ``hot regions''
instead of ``hot spots'' is the correct description of hole doped
cuprate superconductors within the spin-fermion model. Note that since
the maximum of $\left|\chi\right|$ is now at the zone edge, the wave
vector corresponding to the associated charge order is now parallel
to the $x/y$ axes of the system, in similarity with the findings
of Ref.\ {[}\onlinecite{Meier14}{]}.

\subsection{Global trends in cuprate superconductors}

\begin{figure}[tbh]
   \centering    
 \begin{minipage}{4.2cm}
\includegraphics[width=45mm]{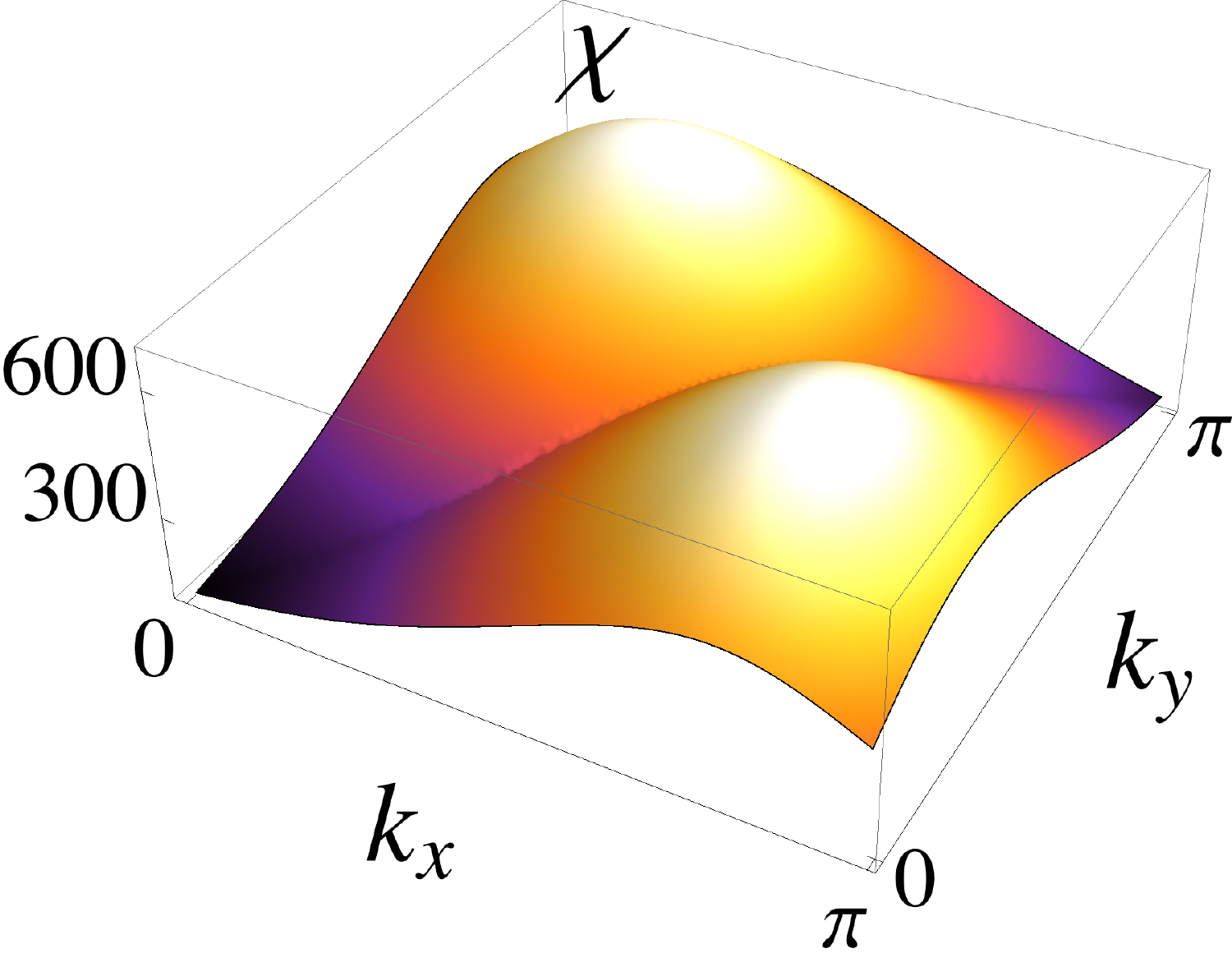}
\end{minipage}
   \begin{minipage}{4.2cm}
\includegraphics[width=45mm]{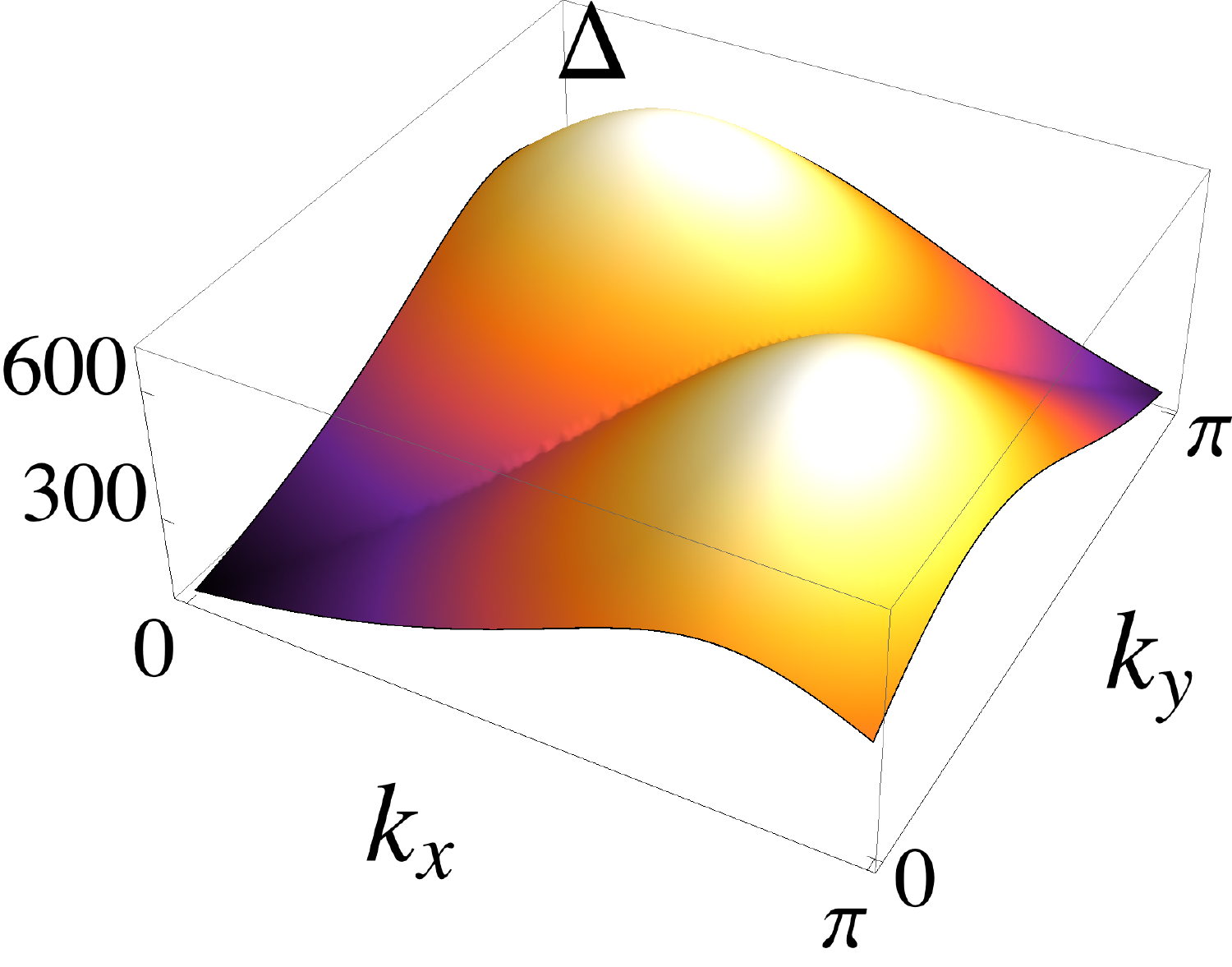}
\end{minipage}
   \begin{minipage}{4.2cm}
\includegraphics[width=36mm]{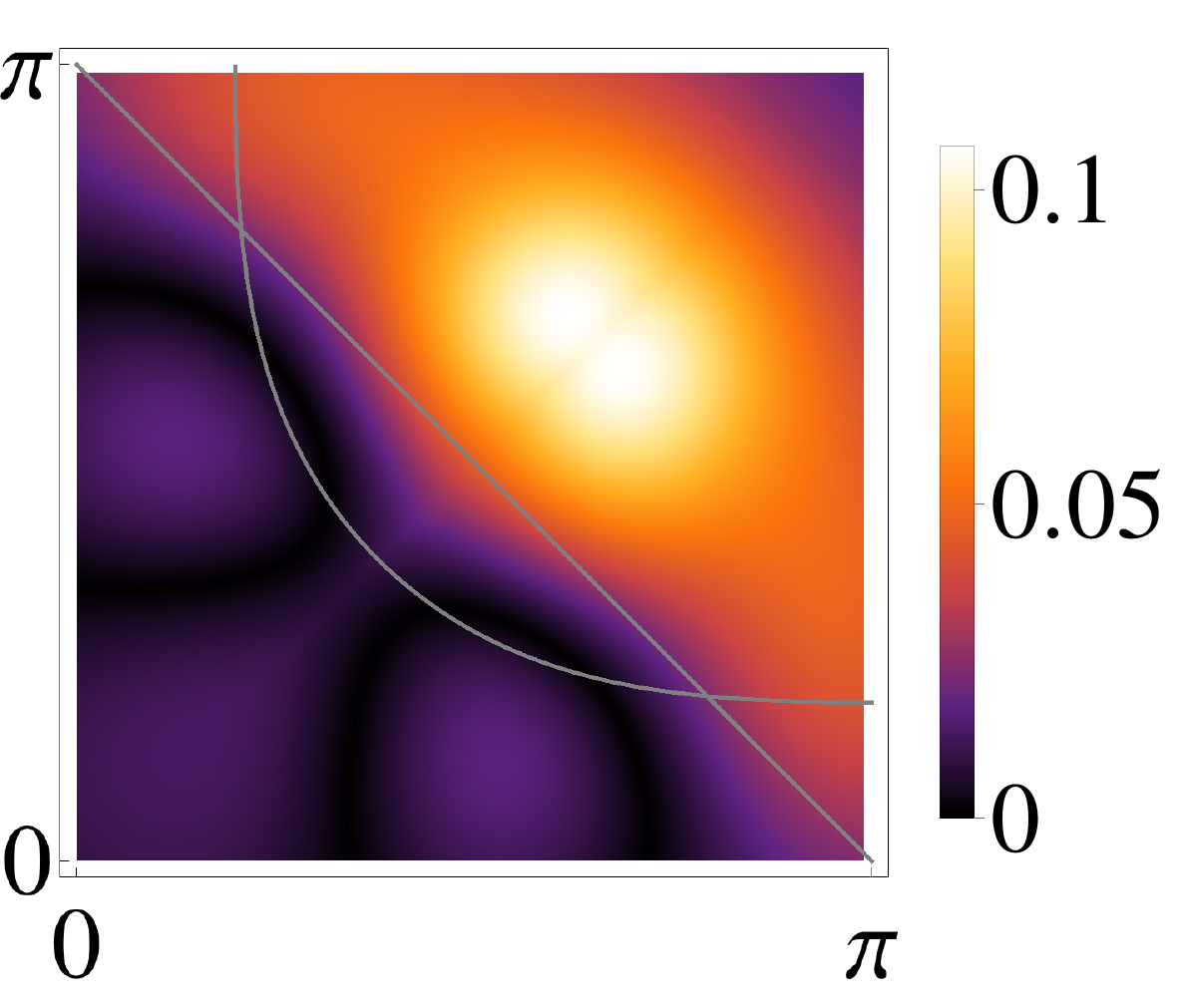}
\end{minipage}
  \vspace{-2mm}
 \caption{\label{fig6} (Color online) Generic picture of the gap functions
$|\chi|$, $|\Delta|$ and $|\chi-\Delta|/|\Delta_{\text{max}}|$
in the first BZ for hole-doped cuprates, here explicitly shown for
YBCO. The figures correspond to the large mass and large coupling
limit ($\lambda=160$ and mass $m=0.5$, so that $E_{\text{0}}\simeq160\,$K
and $E_{\text{eff}}=226\,$K) where the SU(2) symmetry is well respected. }
\end{figure}

\begin{figure}[tbh]
   \centering    
 \begin{minipage}{4.2cm}
\includegraphics[width=45mm]{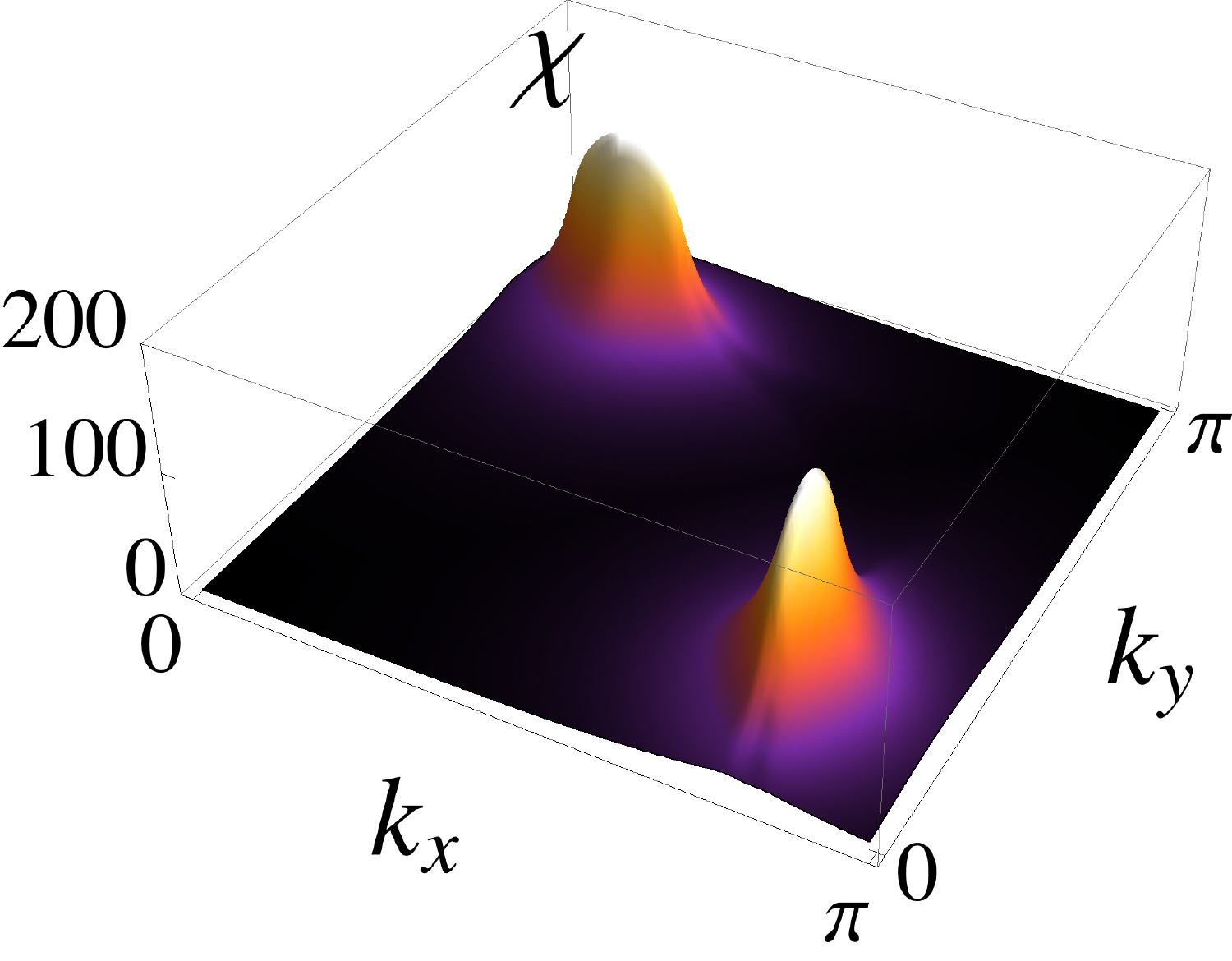}
\end{minipage}
   \begin{minipage}{4.2cm}
\includegraphics[width=45mm]{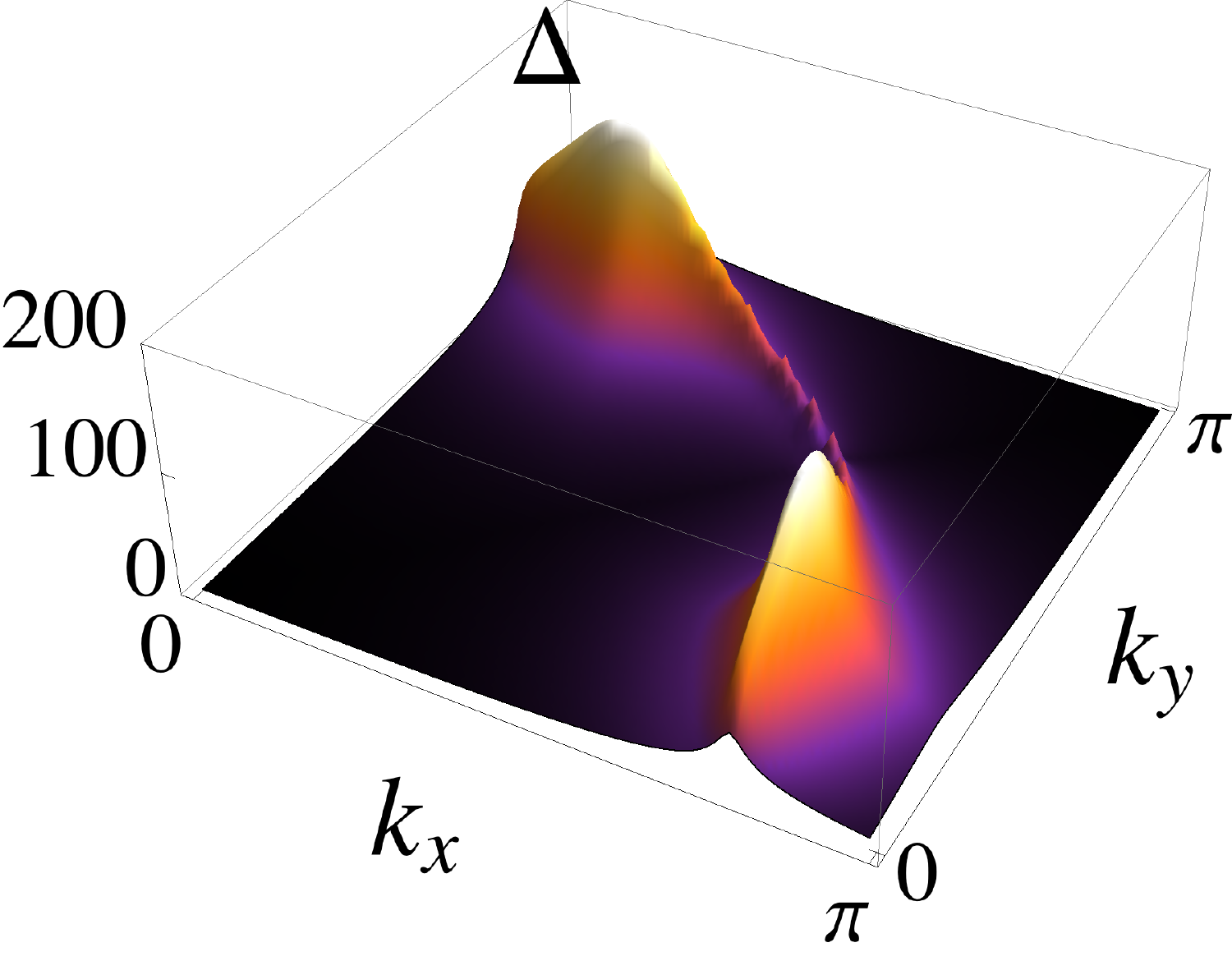}
\end{minipage}
   \begin{minipage}{4.2cm}
\includegraphics[width=36mm]{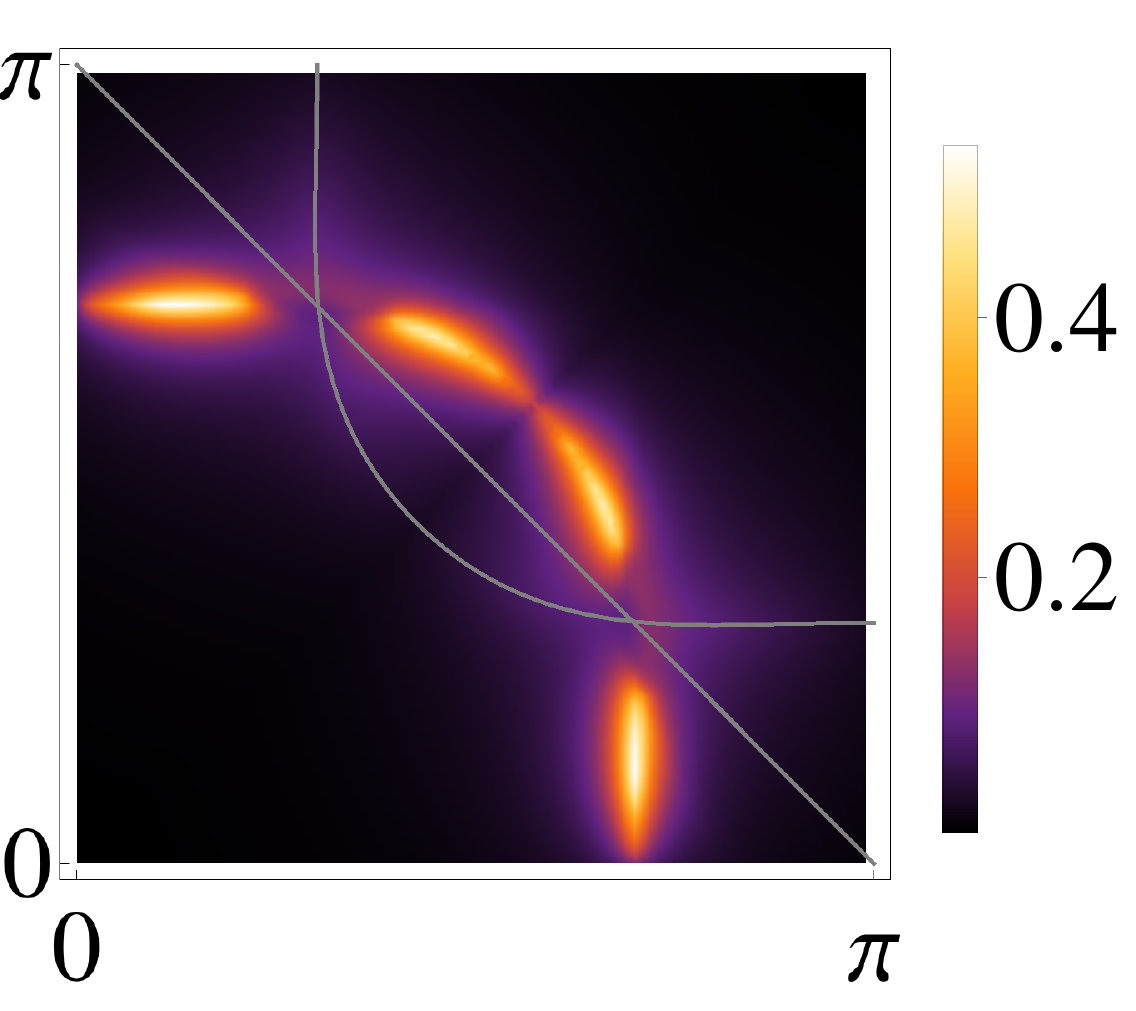}
\end{minipage}
  \vspace{-2mm}
 \caption{\label{fig7} (Color online) Generic picture of the gap functions
$|\chi|$, $|\Delta|$ and $|\chi-\Delta|/|\Delta_{\text{max}}|$
in the first BZ for electron doped materials. The figures correspond
to the small mass and small coupling limit ($\lambda=40$ and mass
$m=10^{-3}$, so that $E_{\text{0}}\simeq40\,$K and $E_{\text{eff}}=1265\,$K)
that is most unfavorable for the SU(2) symmetry. }
\end{figure}

The interplay between mass and coupling allows us to relate global
trends in the phase diagram for cuprate superconductors with the strength
of the SU(2) symmetry breaking. In the electron doped compounds the
coupling between AF modes and conduction electrons is believed to
be weaker than for the hole doped case. From the above analysis we
find that the SU(2) symmetry is less respected in that case leading
to a smaller PG dome and overall smaller SC, see Fig.\ \ref{fig7}.
For the same reasons La compounds where the coupling is also believed
to be small behave similarly. On the other hand, hole-doped cuprates
like YBCO live in the large mass and large coupling regime. This results
in broad gapped regions in the BZ where the symmetry is well respected
so that both the PG and SC dome are large, as shown in Fig.\ \ref{fig6}.

Finally, let us mention that a major effect of a magnetic field is
to invert the order of the level splitting between the CDW and the
SC components \cite{Meier13,Einenkel14}. This will favor the charge
order compared to the SC pairing. We believe that Umklapp scattering
can have a similar effect to invert the level ordering, but leave
detailed investigation for further studies.

\section{Conclusion}

In conclusion, this paper gives firm ground to the intuition that
the charge sector is a key player in the physics of cuprate superconductors.
While the main instability is still the AF ordering, the $d$-wave
bond order relates to the $d$-wave pairing through an SU(2) symmetry.
We have shown that there exists a wide range of parameters where the
SU(2) degeneracy is fulfilled, which gives a natural explanation for
the large PG regime observed in certain compounds. We argue that compounds
like electron doped cuprates or the La compounds are outside the regime
of SU(2) degeneracy, and the more pronounced energy splitting is the
reason for the weaker PG regime.

\section{ACKNOWLEDGMENTS}

We acknowledge discussions with A.\ Chubukov, S.\ Kivelson, H.\ Alloul,
P.\ Bourges, Y.\ Sidis, A.\ Sacuto, and V.\ S.\ de Carvalho.
We thank the KITP, Santa Barbara and the IIP, Natal for hospitality
during the elaboration of this work. This work was supported by LabEx
PALM (ANR-10-LABX-0039-PALM), of the ANR project UNESCOS ANR-14-CE05-0007,
as well as the grant Ph743-12 of the COFECUB which enabled frequent
visits to the IIP, Natal. Numerical calculations were carried out
with the aid of the Computer System of High Performance of the IIP
and UFRN, Natal, Brazil.


\bibliography{Cuprates}

\begin{thebibliography}{42}
\bibitem [{not()}]{note}%
  \BibitemOpen
  \href@noop {} {\ }\bibinfo {note} {Close to a magnetic instability, the
  Spin-Fermion model should capture the relevant low-energy physics of the
  Hubbard model with arbitrary interaction strength, see e.g.\ Ref.\
  [\onlinecite{sfbook}].}\BibitemShut {Stop}%
\end{thebibliography}
 \bibliographystyle{apsrev4-1} 

\end{document}